\documentclass{emulateapj}
\usepackage{times}
\usepackage{epsfig}
\usepackage{graphicx}
\usepackage{latexsym}
\usepackage{pifont}
\input epsf

\shorttitle{{\it Spitzer} IRAC LF of the Coma Cluster}
\shortauthors{Jenkins et al.}

\begin{document}

\title{Uncovering the Near-IR Dwarf Galaxy Population of the Coma Cluster with {\it Spitzer} IRAC}

\author{L. P. Jenkins\altaffilmark{1}, A.E. Hornschemeier\altaffilmark{1}, B. Mobasher\altaffilmark{2}, D.M. Alexander\altaffilmark{3}, F.E. Bauer\altaffilmark{4}}

\altaffiltext{1}{Laboratory for X-ray Astrophysics, NASA Goddard Space Flight Center, Code 662, Greenbelt, MD 20771.}
\altaffiltext{2}{Space Telescope Science Institute, 3700 San Martin Drive, Baltimore, MD 21218.}
\altaffiltext{3}{Royal Society Fellow, Durham University, South Road, Durham, DH1 3LE, UK.}
\altaffiltext{4}{{\it Chandra} Fellow, Columbia University, 2960 Broadway, New York, NY 10027.}

\begin{abstract}

We present the first results of a {\it Spitzer} IRAC (Infrared Array Camera) wide-field survey of the Coma cluster. The observations cover two fields; the first is a 0.733 deg$^2$ region in the core of the cluster (Coma~1), the second a 0.555 deg$^2$ off-center region located $\sim$57\arcmin\ (1.7\,Mpc) south-west from the core (Coma~3). The IRAC observations, although short 70-90\,s exposures, are very sensitive; we detect $\sim$29,200 sources at 3.6$\mu$m over the total $\sim$1.3 deg$^2$ survey area. We construct 3.6$\mu$m galaxy luminosity functions (LFs) for each field using selection functions based on spectroscopic redshifts.  At the bright end, the LFs are well modeled by a traditional Schechter function; $\langle$M$^{\star}_{3.6\mu m}, \alpha_1\rangle = \langle-25.17, -1.18\rangle$ and $\langle-24.69, -1.30\rangle$ for Coma~1 and Coma~3 respectively. However,  at the faint end (M$_{3.6\mu m}> -20.5$), {\it there is a steep increase in the LF slope in both fields indicative of large numbers of red dwarf galaxies}.  The reality of this population is examined using SDSS optical counterparts with optical color filtering ($g-r<1.3$).  The steep increase can be modeled with a power-law function, with slopes of $\alpha_2$ = -2.18 (Coma~1) and $\alpha_2$ = -2.60 (Coma~3), the difference likely indicating a change in environmental effects between the two fields.  A qualitative comparison with optical ($B$- and $R$-band) LFs shows that we are likely to be observing a larger population of dwarf galaxies in the near-IR, which may be a low-surface-brightness (LSB) population that IRAC is particularly sensitive to, or a population too red to be detected in existing optical surveys down to $R\sim$ 20.

\end{abstract}


\section{Introduction}
\label{sec:intro}

The study of the luminosity function (LF) of galaxies in clusters is important to understand the formation and evolution of galaxies.  For example, a testable prediction of standard cold dark matter (CDM) hierarchical clustering models is a steep mass function resulting from the presence of a large number of faint dwarf galaxies, a fossil record of small dark matter halo formation in the early Universe (e.g. \citealt{white78}, \citealt{kauffmann93}).   The LF is defined as the number density of galaxies per unit luminosity, and can be used to directly measure the abundances of various galaxy populations in different environments (e.g. \citealt{trentham02}).  In clusters, it can generally be described by a Schechter function \citep{schechter76}, with the bright end (M$_{R} \la$ -18) dominated by giant elliptical galaxies, and the faint end (M$_{R} \ga$ -18) dominated by red dwarf elliptical (dE) galaxies (e.g. \citealt{secker96}; \citealt{secker97}; \citealt{trentham98};  \citealt{mobasher03}).  Observations of nearby clusters allow us to probe the faint end of the LF in detail, and there is now substantial evidence that clusters contain significantly higher numbers of low-luminosity dwarf galaxies than are found in the field (e.g. \citealt{trentham05}; \citealt{popesso06}; \citealt{adami06}).  

Although it seems that the most massive galaxies in clusters formed at early times ($z > 2$), it is not known whether dwarf galaxies also formed early-on within the cluster or whether they were accreted from the field at lower redshifts ($z < 0.5$).  There is, however, growing evidence from infrared (IR) LF studies in support of the latter scenario (e.g. \citealt{depropris99}; \citealt{muzzin05}).  This is further supported by studies of the spatial distribution and radial velocities of dEs in the Virgo cluster \citep{conselice01}, where it was found that the cluster dwarf population probably originated as field spiral galaxies that were accreted and transformed into dEs by the process of `galaxy harassment' \citep{moore98}.   We may therefore expect to see measurable differences between the dwarf populations in outer regions of clusters (where field galaxies are accreted) and the denser central regions, where destructive environmental factors also come into play (see \S~\ref{sec:environ}).  Faint populations attached to infalling galaxy groups may also increase dwarf galaxy densities in clusters, although they would be more susceptible to tidal stripping than brighter group galaxies and may therefore be expected to be located preferentially away from the core of the cluster \citep{biviano96}.  It is possible to directly search for such environmental differences by observing different regions within a cluster and constructing region-specific LFs in a consistent fashion.

At a redshift of $z\sim0.023$, Coma (Abell~1656) is one of the most extensively studied clusters (see \citealt{biviano98} for a historical review), and its richness and proximity make it an ideal target for the study of the faint-end LF.  It is approximately spherically symmetric, and has two central dominant elliptical galaxies (NGC~4874 \& NGC~4889).  Although once thought to be fully relaxed, several sub-structures have been found at optical and X-ray wavelengths (e.g. \citealt{mellier88}, \citealt{white93}, \citealt{neumann03}).  One is centered on the central galaxy NGC~4874, another on a sub-group of galaxies around the giant elliptical NGC~4839, located 40\arcmin\ south-west of the center.   Recent X-ray observations with {\it XMM-Newton} have shown the NGC~4839 sub-group to be falling into the cluster core \citep{neumann01}. 

An accurate determination of the faint-end slope of the LF of the Coma cluster is an essential step towards constraining the mass function of galaxies in the local Universe.  Numerous studies, mainly at optical wavelengths, have measured steep faint-end slopes ($\alpha\sim -1.7$ to $-2.3$, e.g. \citealt{trentham98}; \citealt{milne07}; \citealt{adami07}), while others see a fairly flat distribution (e.g. $\alpha\sim-1.2$,  \citealt{mobasher03}).  These discrepancies are likely, however, to be due to both the different photometric limits of the surveys and the different methods of subtracting the background galaxy component (see \S~\ref{sec:compare_other}).  

In this paper we aim to improve the determination of the dwarf galaxy population of Coma by measuring its near-IR (3.6$\mu$m) LF in two regions of different densities with the Infrared Array Camera (IRAC, \citealt{fazio04a}) on board the {\it Spitzer} Space Telescope.  Observations in the near-IR provide us with a different perspective on the cluster LF to that at shorter wavelengths, as we observe emission from old stellar populations as opposed to the younger populations detected in the optical.  The near-IR is therefore less sensitive to star-formation in galaxies and provides a more direct measure of the underlying galaxy mass function of the cluster.   Also, near-IR measurements are relatively free of biases due to dust obscuration, which means we have the potential to detect low-luminosity dusty galaxies that could be missed in optical surveys.  In the past, the small field-of-view (FOV) of ground-based near-IR detectors has made extensive surveys of nearby clusters difficult.  However, IRAC efficiently mosaics large areas of the sky, and has both sufficient spatial resolution ($\sim2$\arcsec) {\it and} sensitivity to perform such a survey.  In the shortest wavelength IRAC band (3.6$\mu$m), the IR zodiacal light is at a minimum of  $\sim$23.5 mag arcsec$^{-2}$ \citep{kelsall98},  a factor of 10,000 lower than ground-based $K$-band imaging ($\sim$13.5 mag arcsec$^{-2}$, \citealt{wainscoat92a}).  The sensitivity and spatial resolution of the IRAC 3.6$\mu$m observations, together with a redshift-based statistical background source subtraction, foreground star subtraction based on Galaxy star counts, and conservative optical color filters to guard against background galaxies, will allow us to make a better determination of  the size of the dwarf galaxy population of Coma.  A well established 3.6$\mu$m LF at $z\sim0$ can then be compared with rest-frame 3.6$\mu$m LFs at higher redshifts derived from other IRAC channels.  For example, 8$\mu$m IRAC observations sample 3.6$\mu$m emission at $z\sim1$, which means we could study the equivalent near-IR LF (i.e. the mass function of galaxies) at this redshift to see how massive clusters evolve. 

In \S~\ref{sec:obs} we describe the {\it Spitzer} IRAC observations and data reduction procedures.  In \S~\ref{sec:sources} we describe our source detection and photometry, and methods used to separate stars from galaxies.  In \S~\ref{sec:optical_survey} we outline the optical photometric/spectroscopic surveys we use to identify Coma member galaxies, and in \S~\ref{sec:members} we determine what fraction of IRAC detected galaxies are members of Coma.  In \S~\ref{sec:lf} we construct IRAC LFs for both the core and off-center fields.  The results are discussed in \S~\ref{sec:disc} and compared with those found at other wavelengths.  Conclusions are given in \S~\ref{sec:conc}. Throughout this paper we use a distance modulus (DM) of $m-M=35.0$ mag, corresponding to a distance to Coma of 100\,Mpc for $H_0$=70\,km s$^{-1}$ Mpc$^{-1}$.

\begin{figure}
\begin{center}
\scalebox{0.45}{\includegraphics{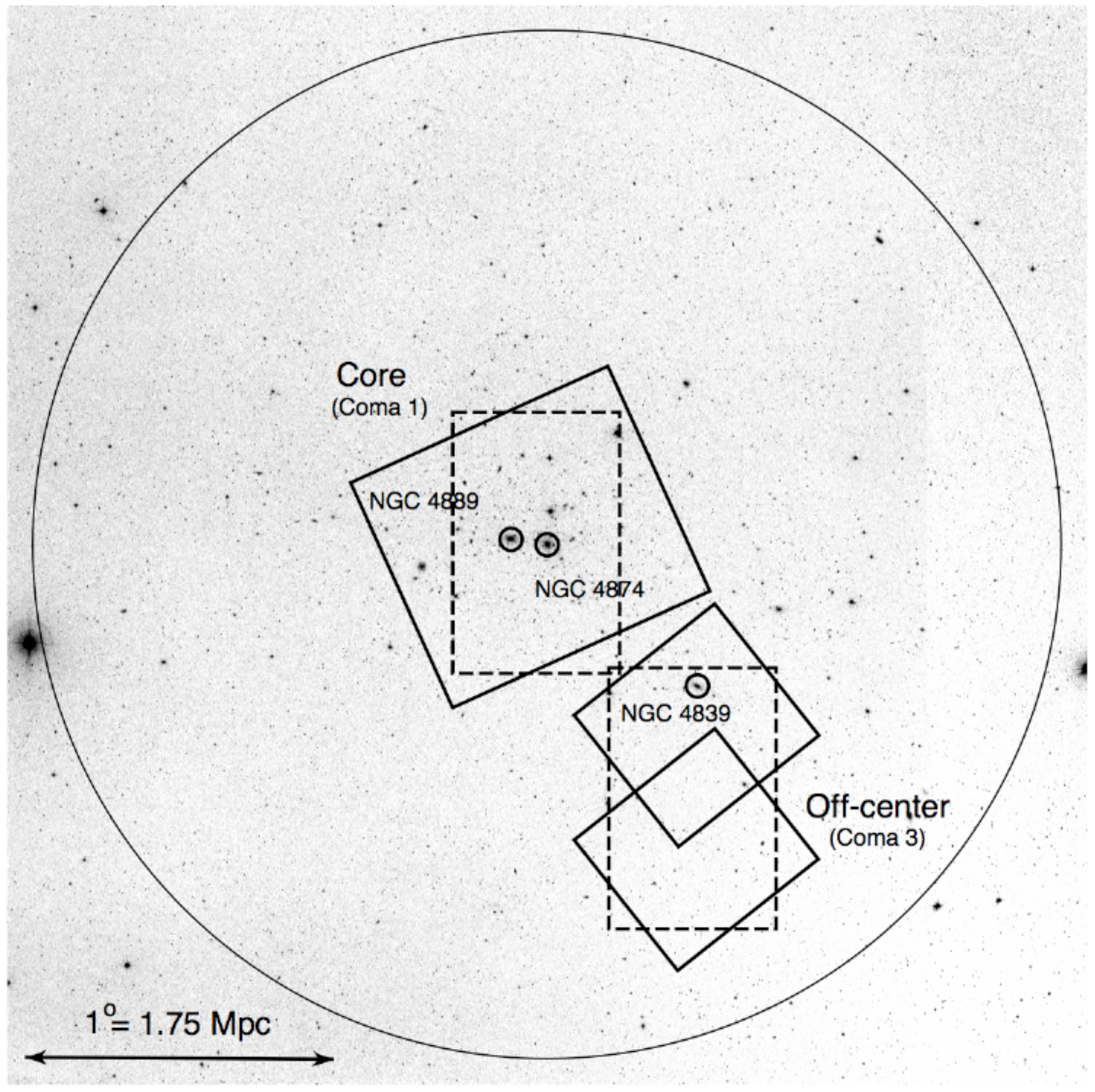}}
\caption{DSS1 image of the Coma cluster showing the location of the two {\it Spitzer} IRAC fields ({\it solid boxes}).   The optical spectroscopic survey fields of \cite{mobasher01} are shown with dashed boxes, and the circle denotes the virial radius of the cluster (100\arcmin or 2.9\,Mpc, \citealt{lokas03}).  The three giant elliptical galaxies are also marked; NGC~4889, NGC~4874 \& NGC~4839. There are 420 optically identified member galaxies in these two fields.}
\label{fig:dss}
\end{center}
\end{figure}

\section{Observations and Data Reduction}
\label{sec:obs}

The Coma cluster is approximately spherically symmetric, with a virial radius of 100\arcmin\ (2.9\,Mpc) and total mass,  estimated from dynamical modeling, of 1.4$\times10^{15}h_{70}^{-1}$M$_{\odot}$ \citep{lokas03}.  The aim of this study is to characterize the near-IR dwarf galaxy population of Coma, and to investigate whether the LF changes, due to e.g. accretion of field dwarf galaxies/environmental factors, between regions of different cluster-centric radius and density.   We select two regions of Coma to overlap with the spectroscopic survey performed by \cite{mobasher01}; these are shown in Figure~\ref{fig:dss}.  The optical fields are each $\sim32\farcm5\times$50\farcm8; one is located at the core (Coma 1, centered at 12$^{h}$59$^{m}$23\fs7, +28\degr01\arcmin12\farcs5), and the second south-west of the core (Coma 3, centered at 12$^{h}$57$^{m}$07\fs5, +27\degr11\arcmin13\farcs0).  These two fields were chosen firstly because of their large density contrasts (the local density of Coma 3 is somewhere between Coma 1 and the field), and secondly because the Coma 3 field covers the X-ray secondary peak (the giant elliptical galaxy NGC~4839).  We refer to Coma 1 and Coma 3 as the `core' and `off-center' fields respectively.  The center of the Coma 3 field is located $\sim$57\arcmin\ (1.7\,Mpc) from the established central galaxy NGC~4874 \citep{kent82}, i.e. $\sim$0.6 times the virial radius.

\subsection{{\it Spitzer} IRAC Observations}
\label{sec:irac_obs}

The Coma 3 off-center field was observed by {\it Spitzer} on June 13, 2005 (Program ID 3521). The observation consists of two pointings designed to cover as much of the Coma 3 field as possible.  AOR 11067904 (Top) is centered on 12$^{h}$57$^{m}$25\fs0, +27\degr22\arcmin13\farcs0, while AOR 11067648 (Bottom) is centered on 12$^{h}$57$^{m}$25\fs0, +26\degr58\arcmin13\farcs0, the two fields overlapping in the middle.  Each pointing is made up of an 8$\times$9 grid of 72\,s per pixel exposures, dithered over 6 positions with a medium scale cycling pattern to fully sample the point spread function (PSF). Since there are a few moderately bright sources in the field, the high dynamic range full array mode was used, which includes some very short exposures (0.6\,s) to aid photometry and PSF fitting of saturated sources.  Note that these short exposures are not used when combining the images (see \S~\ref{sec:irac_images}), as this would result in higher background noise and decreased sensitivity.

To study the Coma 1 core field, we use archival data taken on the core of the Coma cluster.  These data, gathered by the IRAC instrument team in July 2004 (Program ID 25, P.I. Fazio), are centered on 12$^{h}$59$^{m}$48\fs7, +27\degr58\arcmin50\farcs0 and are slightly deeper than the Coma 3 off-center data (90\,s per pixel).  This observation is comprised of two 12$\times$6 grids (AORs 3859712 \& 3859968) taken in standard full array mode, dithered over 3 positions with a small cycling pattern.

\subsection{IRAC Data Reduction}
\label{sec:irac_images}

\begin{figure}
\begin{center}
\scalebox{0.45}{\includegraphics{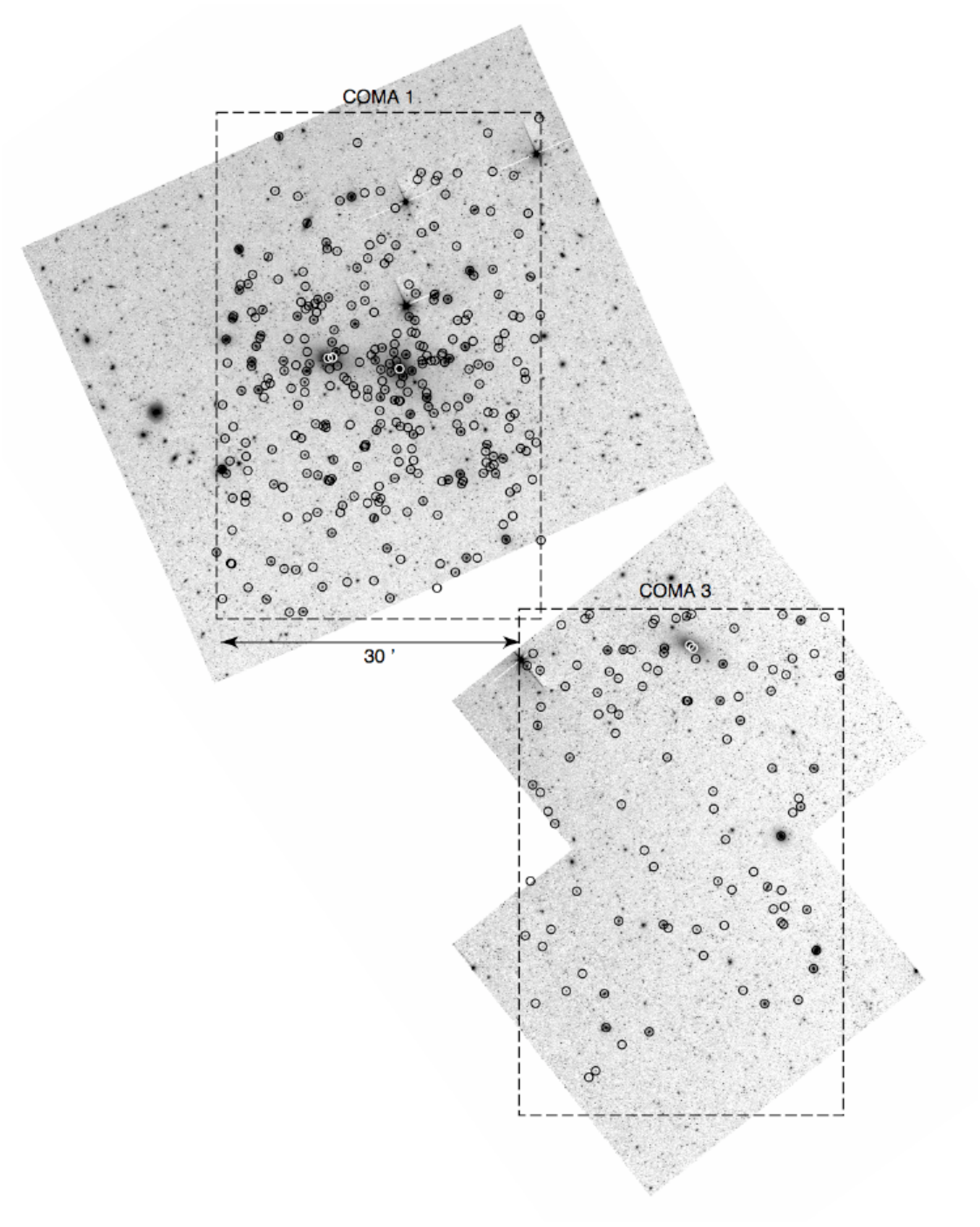}}
\caption{The 3.6$\mu$m {\it Spitzer} IRAC mosaic of the core (Coma 1, 0.733 deg$^2$) and off-center (Coma 3, 0.555 deg$^2$) regions of the Coma cluster.  The fields of the optical photometric/spectroscopic survey of \cite{mobasher01} are marked with rectangles, each 32.5$\times$50.8\arcmin.  The circles denote the 410 spectroscopically confirmed members of the cluster detected above 3$\sigma$ in the IRAC data. }
\label{fig:irac_mosaic}
\end{center}
\end{figure}

The Basic Calibrated Data (BCD) were supplied by the {\it Spitzer Science Center} (SSC), consisting of individual calibrated frames. The pipeline processing steps performed to produce these data were dark subtraction, flat fielding and flux calibration (see the IRAC Data Handbook for further details).  We performed post-BCD processing on the GO and archival data in the following manner. The individual images were mosaiced together using the SSC's data processing package, {\tt MOPEX}, specifying a standard pixel size of 1.22\arcsec. In order to exactly align the images of the 4 IRAC bands, Fiducial Image Frames (FIFs) were created for each observation using all BCD images in that observation and applied to the mosaicing of each individual band image. A full mosaic of the two fields is shown in Figure~\ref{fig:irac_mosaic}.

The images are flux calibrated in the SSC pipeline in surface brightness units (MJy~sr$^{-1}$). For the purposes of determining individual source fluxes (see \S~\ref{sec:sourcedetect}), the images were converted to flux density units by multiplying by a conversion factor of 1MJy~sr$^{-1}$ = 34.98\,$\mu$Jy per pixel.   This is a combination of the conversion given in the {\it Spitzer Space Telescope} Observers Manual (1MJy~sr$^{-1}$ = 23.50443 $\mu$Jy arcsec$^{-2}$), and the pixel solid angle (1.22 arcsec$^2$, since the flux density of an object is the integral of the surface brightness over its solid angle).

\section{Source Detection, Photometry \& Star-Galaxy Separation}
\label{sec:sources}

\subsection{IRAC Source Detection \& Photometry}
\label{sec:sourcedetect}

Source detection was carried out with the SExtractor (SE) algorithm \citep{bertin96}.  The SE detection threshold was set to 3$\sigma$, with the {\tt SEEING\_FWHM} parameter fixed to 1.7\arcsec\ to match the IRAC 3.6$\mu$m PSF full-width-at-half-maximum (FWHM, see table~3, \citealt{fazio04a}). Only sources within the areas observed in all four IRAC bands were included, to be able to calculate IRAC colors for every source.  We used the `double-image' mode in SE, detecting objects in the image with the highest signal-to-noise (3.6$\mu$m).  This provides photometry on 4 IRAC bands, and allows a consistent measure of IRAC colors. The total areas analyzed in each field were 0.733 deg$^2$ (Coma 1)  and 0.555 deg$^2$ (Coma 3). 

We measured `total' fluxes using flexible elliptical apertures (i.e. {\tt FLUX\_AUTO} in SE), whose elongation and position angle are defined by the object's light distribution, and therefore proportional to the size of each galaxy in the image  \citep{bertin96}.  The {\it Spitzer} SWIRE team has found that this is the best approach for measuring the flux from small (a few arcsecond size) extended sources (see section 5.8 in the IRAC Data Handbook v.3). The {\tt AUTO} elliptical apertures were set to 2.5 times the characteristic `Kron' radius of each source, so that they contain $\ga$ 95\% of source flux \citep{kron80}. Since the Coma field is dense, the local background was subtracted using a 5 pixel radius annulus around each source to mitigate contamination from nearby sources. The measured flux densities were then converted to apparent magnitudes using the flux density zero points (relative to Vega) quoted in \cite{reach05}; these are 280.9\,Jy (3.6$\mu$m), 179.7\,Jy (4.5$\mu$m), 115.0\,Jy (5.8$\mu$m) and 64.13\,Jy (8$\mu$m).  Finally, the images and source detections were visually inspected.  Spurious sources (e.g. those in the trails of very bright stars) were removed.

\begin{figure}
\begin{center}
\scalebox{0.35}{\includegraphics{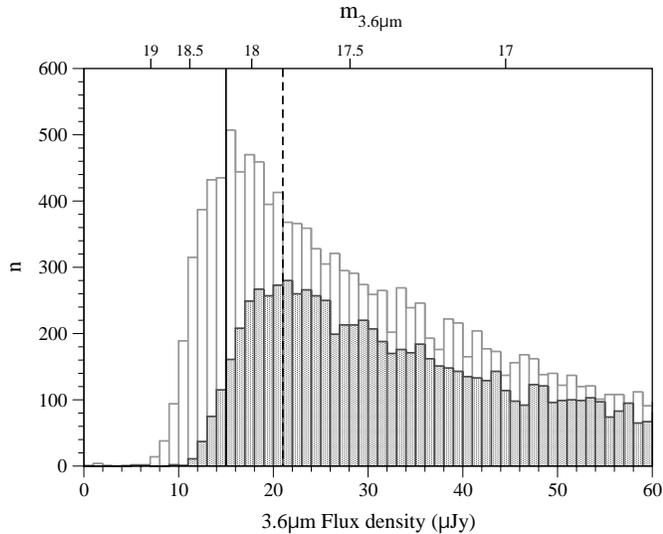}}
\caption{The 3.6$\mu$m flux density distribution of sources detected in the IRAC data.  The open histogram represents the distribution of the 17,872 sources detected in the Coma 1 core field, and the shaded histogram shows the same for the 11336 sources in the Coma 3 off-center field.  The solid vertical and dashed lines show where major incompleteness starts to affect the Coma 1 data (15\,$\mu$Jy, m$_{3.6\mu m}\sim$18.2) and Coma 3 data (21\,$\mu$Jy, m$_{3.6\mu m}\sim$17.8) respectively.  To guard against any minor incompleteness effects brighter than this, we adopt more conservative limits of m$_{3.6\mu m}=17.0$ for Coma 1 and m$_{3.6\mu m}=16.5$ for Coma 3 for the LF calculations.}
\label{fig:sensitivity}
\end{center}
\end{figure}

The IRAC catalog, which will be published separately, contains four-band data for a total of 29,208 sources, 17,872 in Coma 1 and 11,336 in Coma 3.  Figure~\ref{fig:sensitivity} shows the flux density distribution of the sources detected in both fields. These turn over due to the onset of incompleteness at $\sim15\mu$Jy  (m$\sim$18.2) in  the Coma 1 core field and $\sim21\mu$Jy (m$\sim$17.8) in the Coma 3 off-center field at 3.6$\mu$m (shown with vertical lines).   However, to guard against any minor effects of incompleteness at brighter magnitudes, we adopt more conservative completeness limits of m$_{3.6\mu m}=17.0$ and 16.5 (corresponding to M$_{3.6\mu m}=$ -18 and -18.5) for Coma 1 and 3 respectively for our LF calculations (see \S~\ref{sec:lf}).

\begin{figure}
\begin{center}
\scalebox{0.48}{\includegraphics{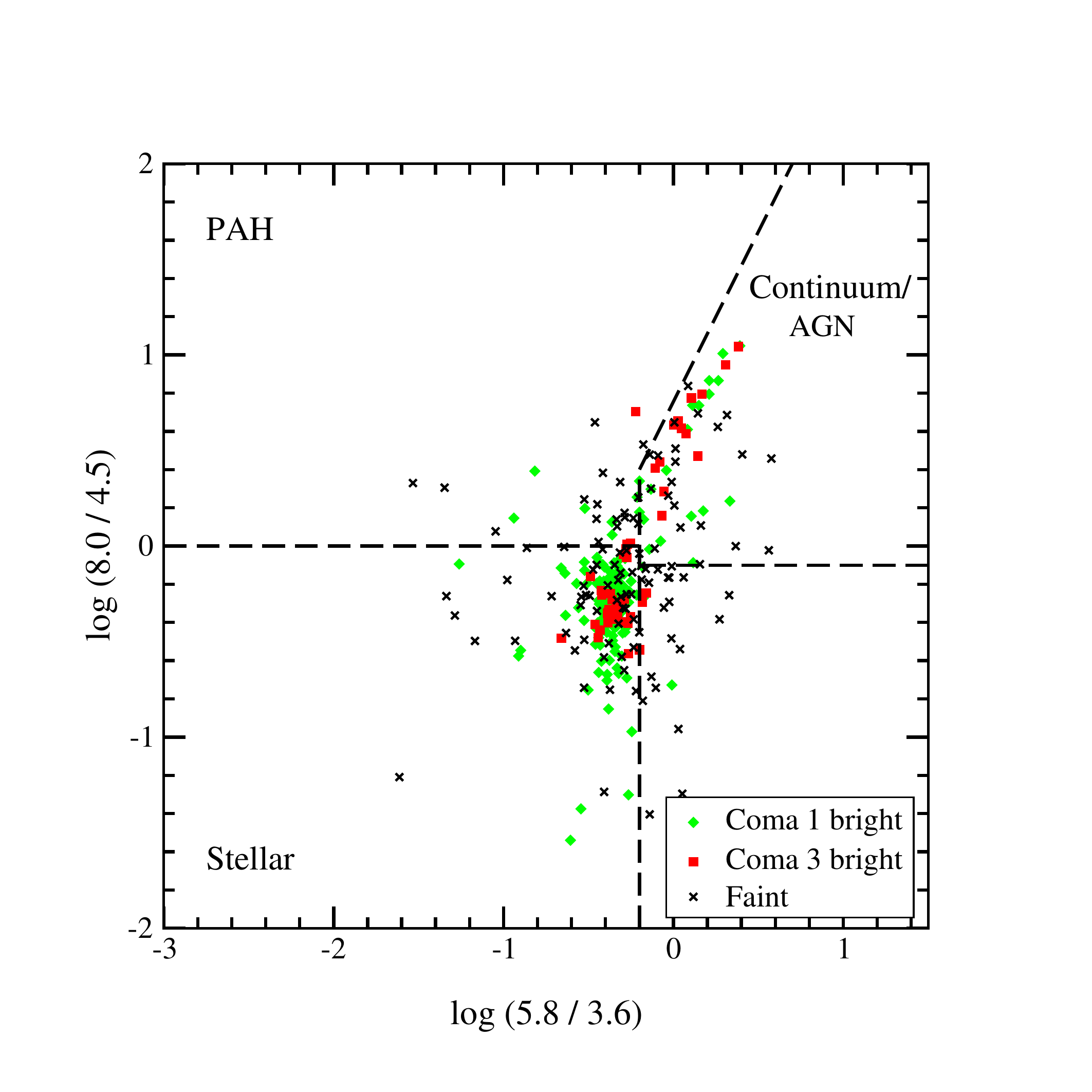}}
\caption{IRAC color-color diagram of the confirmed Coma members detected in all 4 bands in both fields, defined in terms of the logarithms of flux densities in the 5.8/3.6$\mu$m and 8.0/4.5$\mu$m bands.  The dashed lines mark boundaries between different galaxy types as defined by \citet{sajina05}. Green diamonds and red squares denote the bright (m$_{3.6\mu m} <14.5$) spectroscopic members in Coma 1 and 3 respectively, while the fainter members are shown with crosses.  The majority of the bright Coma members have colors consistent with passive stellar galaxies.}
\label{fig:irac_colors}
\end{center}
\end{figure}

To give an insight into the different galaxy populations in Coma, Figure~\ref{fig:irac_colors} shows an IRAC color-color diagram  (5.8/3.6$\mu$m vs 8.0/4.5$\mu$m) for the spectroscopically confirmed bright (m$_{3.6\mu m} <14.5$)  and faint ($14.5 <$ m$_{3.6\mu m} < 18$)  members in the two fields (see \S~\ref{sec:members} for member identification).  This combination of colors is a good indicator of star-formation activity in galaxies (i.e. ratios of PAH features to stellar light).  The four areas denote the expected colors for PAH-dominated (star-forming), stellar-dominated (passive) and continuum-dominated (active; AGN) galaxies in the four IRAC bands at zero redshift, as predicted by \cite{sajina05}.  This shows that the majority of the Coma members have colors consistent with those of passive stellar galaxies, as expected for such a massive cluster.  Another population have colors consistent with AGN, which can be confirmed by follow-up X-ray observations with {\it Chandra} and {\it XMM-Newton}.  Note that the bright members are skewed towards the passive galaxy colors, while the faint members randomly populate the color-color space.

\subsection{Star-Galaxy Separation}
\label{sec:stars}

According to the IRAC source number counts of  \cite{fazio04b}, a substantial fraction of sources in the IRAC bands, particularly at bright magnitudes, are foreground Galactic stars.  In order to reliably measure the number of galaxies in the data, it is important to filter these out.  At the brightest magnitudes in our catalog (m$_{3.6\mu m}<14.5$), we can use the spatial extent of galaxies at 3.6$\mu$m to morphologically separate them from stars. Here we use the method described in \cite{fazio04b} and \cite{eisenhardt04}, which uses the difference between small 3\arcsec\ diameter circular aperture and `auto' aperture flux measurements in SE as a `concentration parameter'.   In this scheme, which is reliable down to m$_{3.6\mu m}\sim$14.5 \citep{fazio04b}, point sources will have approximately the same magnitude using both `auto' and circular (corrected) apertures, whereas galaxies will show an excess in the corrected circular aperture because their emission does not match the IRAC PSF profile.

\begin{figure}
\begin{center}
\rotatebox{270}{\scalebox{0.35}{\includegraphics{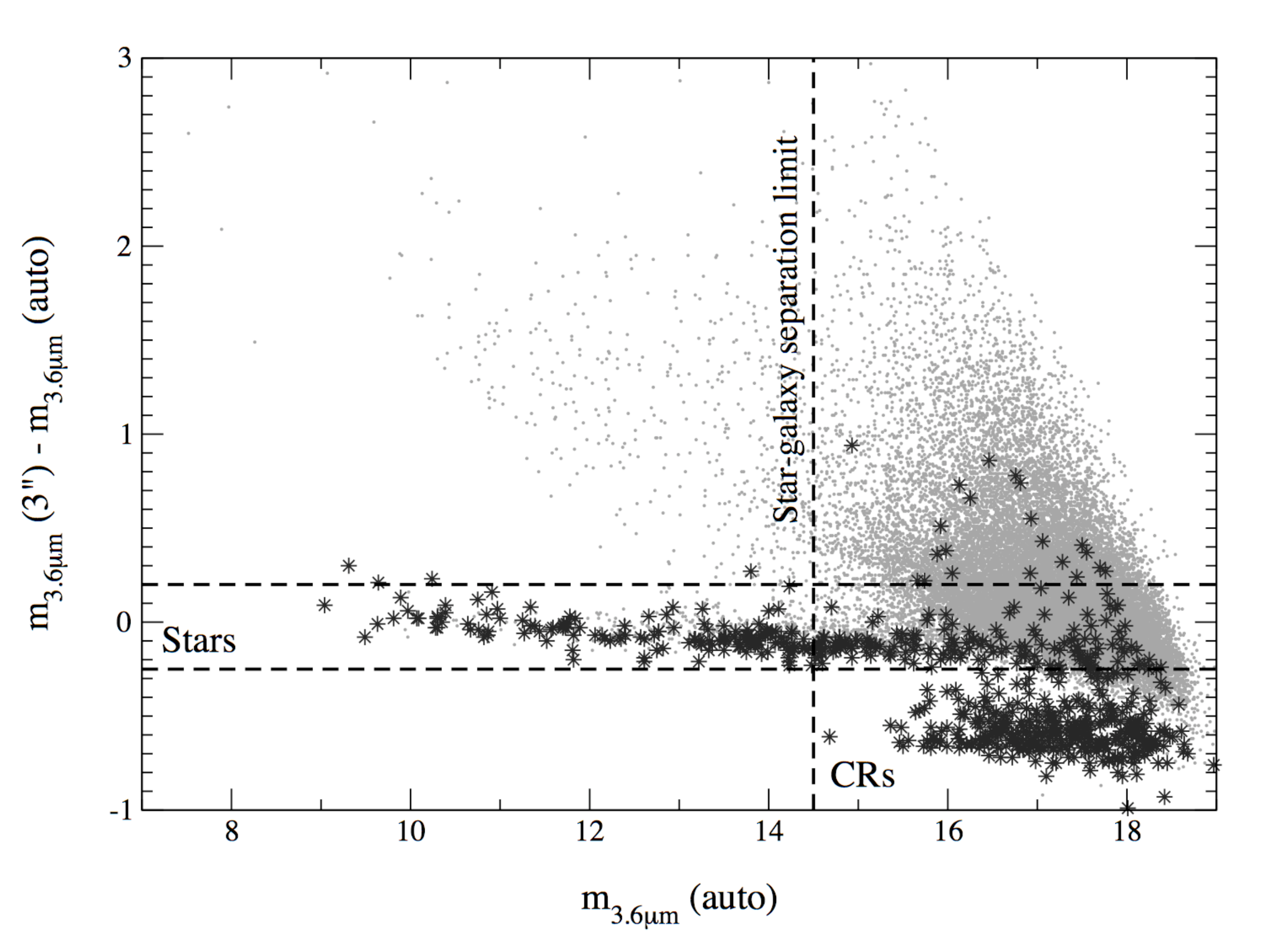}}}
\caption{The stellar concentration parameter, as defined by \cite{fazio04b} and \cite{eisenhardt04} (see text for details).  All sources detected in both Coma fields are marked with small points, and those with the SE {\tt CLASS\_STAR} parameter $>$ 0.95 are over-plotted with stars. The horizontal lines denote the boundaries of the stellar classification, and the vertical line shows the magnitude down to which this is reliable (m$_{3.6\mu m}=14.5$).  Sources below the lower bound of the stellar classification {\it and} fainter than m$_{3.6\mu m}=14.5$ (lower right-hand corner) are classed as cosmic rays (CRs).  All sources brighter than m$_{3.6\mu m}=14.5$ classed as stars, plus fainter sources classed as CRs, are excluded from the LF calculations (see text for further details).  Fainter than this, stars are subtracted from the LFs using the DIRBE Faint Source Model (see Figure~\ref{fig:starcounts}).}
\label{fig:starconc}
\end{center}
\end{figure}

\begin{figure}
\begin{center}
\scalebox{0.45}{\includegraphics{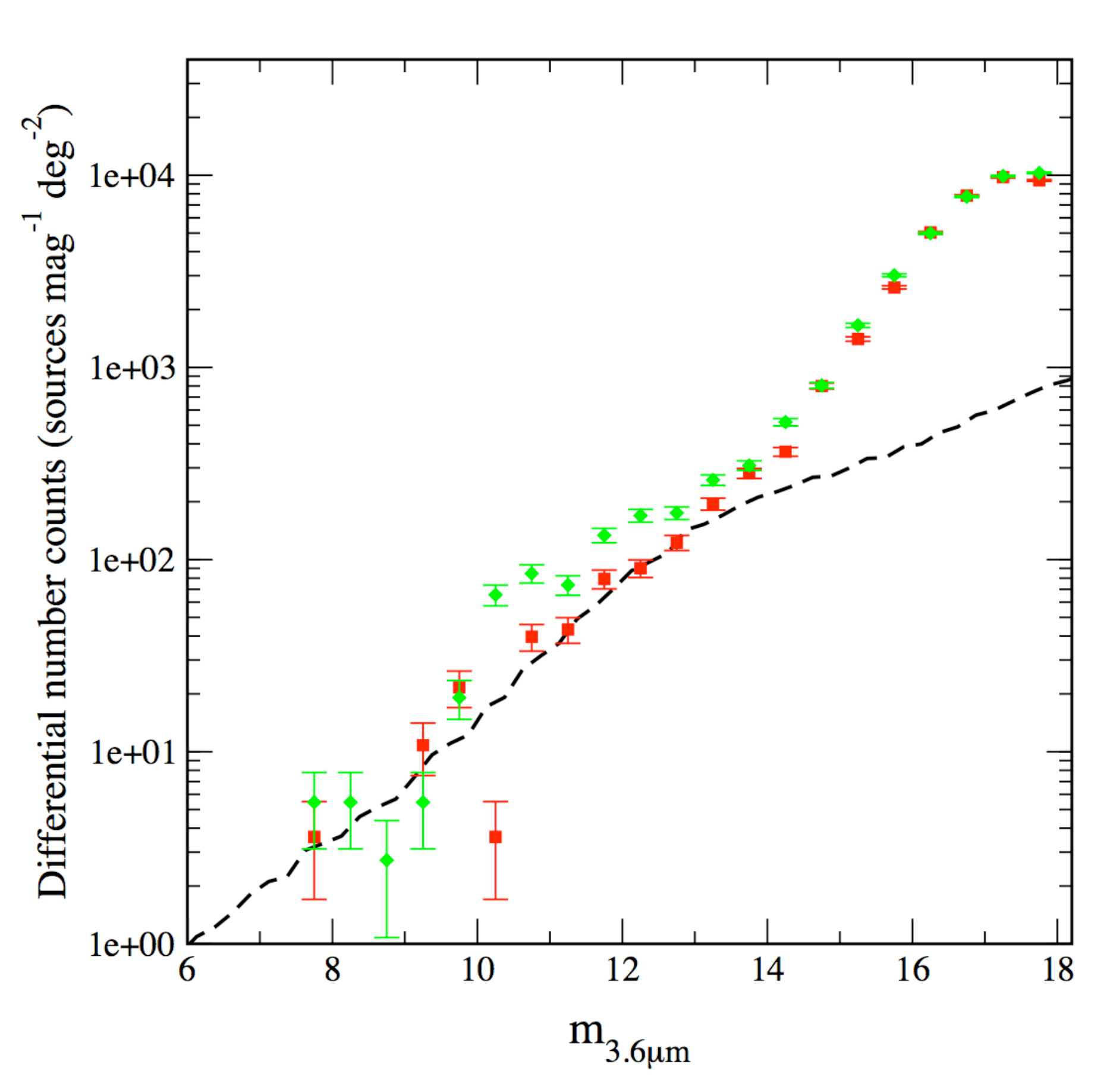}}
\caption{Differential number counts of all sources detected at 3.6$\mu$m in the Coma 1 core field ({\it green diamonds}) the and Coma 3 off-center field ({\it red squares}).  The predicted number counts of Galactic stars from the DIRBE Faint Source Model \citep{arendt98} for the line-of-sight towards Coma (Galactic coordinates: $l=58.08, b=87.96$) are shown as a dashed line.  At bright magnitudes (m$_{3.6\mu m}<14.5$), stars are subtracted from the LF calculations using the stellar concentration parameter (Figure~\ref{fig:starconc}).  Fainter than this, the stellar contribution is statistically subtracted using the Faint Source Model. }
\label{fig:starcounts}
\end{center}
\end{figure}

Figure~\ref{fig:starconc} illustrates the concentration parameter, showing the boundaries defined by \cite{eisenhardt04} to separate stars from galaxies. Sources with values of the SE parameter {\tt CLASS\_STAR} $>$ 0.95 are also shown (where values close to 1 represent point sources), demonstrating that these sources do indeed fall inside the designated ranges.  An aperture correction of -0.69 mag was derived from the data to correct the 3\arcsec\ aperture magnitudes out to the IRAC total magnitude calibration diameter of 24.4\arcsec~(20 pixels).  This is defined, in terms of flux, as the average ratio (1.90) of the 24.4\arcsec\ to 3\arcsec\ aperture measurements from SE on 56 unsaturated stars (9 $<$ m$_{3.6\mu m}<$ 12) using a 5 pixel-wide background annulus (this magnitude range excludes the few brightest stars that are obviously saturated).   Note that this small aperture size does not provide a reliable measurement of the total fluxes of galaxies, but it is useful for the purpose of star-galaxy separation.  All sources satisfying the criterion -0.25 $<$ m$_{3.6\mu m}$ (3\arcsec)  -- m$_{3.6\mu m}$ (auto) $<$ 0.2 and brighter than 14.5  mag were flagged as possible stars.  These were all inspected visually, and 7 were re-classified as galaxies based on their IRAC morphology.  As an additional check, the positions of the sources classified as stars were cross-matched with the optical photometric catalog of \cite{komiyama02} (which has also had the stars removed by a similar method) using a 3$\sigma$ matching radius of 4.24\arcsec~(see \S~\ref{sec:members}), and only 4 sources were reclassified as galaxies based on these matches.  The 4 brightest stars in these fields were also classified as galaxies by their concentration parameter due to the bright wings of their PSFs, and these were subsequently re-classified as stars after visual inspection of the data.  In total, 476 bright sources (m$_{3.6\mu m}<14.5$) were classified as stars over the two fields, and removed from the LF calculations.  We also removed 2100 sources with m$_{3.6\mu m}$ (3\arcsec)  -- m$_{3.6\mu m}$ (auto) $<$-0.25 fainter than 14.5 mag as these are almost certainly cosmic rays \citep{eisenhardt04}.  

At fainter magnitudes (m$_{3.6\mu m}\ga$ 14.5), it is difficult to distinguish stars from galaxies morphologically, so we must subtract the stellar contribution using statistical methods.  Here we use the ``DIRBE (Diffuse Infrared Background Experiment) Faint Source Model" for Galactic stars, which is an implementation of the \cite{wainscoat92b} statistical model (further details can be found in \citealt{arendt98}).  Figure~\ref{fig:starcounts} shows the differential number counts of the two Coma datasets, together with the predicted model counts for Galactic stars in the DIRBE 3.5$\mu$m band for the line-of-sight towards Coma (Galactic coordinates: $l=58.08, b=87.96$, R. Arendt, {\it private communication}). The model demonstrates the dominance of stars at bright magnitudes (8 $<$ m$_{3.6\mu m}<$ 14). However, the counts in the central Coma 1 field are well above the star counts from as bright as m$_{3.6\mu m}\sim10$ due to the high number of bright galaxies located there, whereas the counts in the off-center field do not diverge significantly from the model until m$_{3.6\mu m}\sim13$.   Fainter than m$_{3.6\mu m}\sim14.5$, there is a much smaller stellar contribution to the total number counts ($\sim$10--15\%), and we subtract the model star counts from our data in this range for our LF calculations in \S~\ref{sec:lf}.

\begin{figure}
\begin{center}
\rotatebox{90}{\scalebox{0.35}{\includegraphics{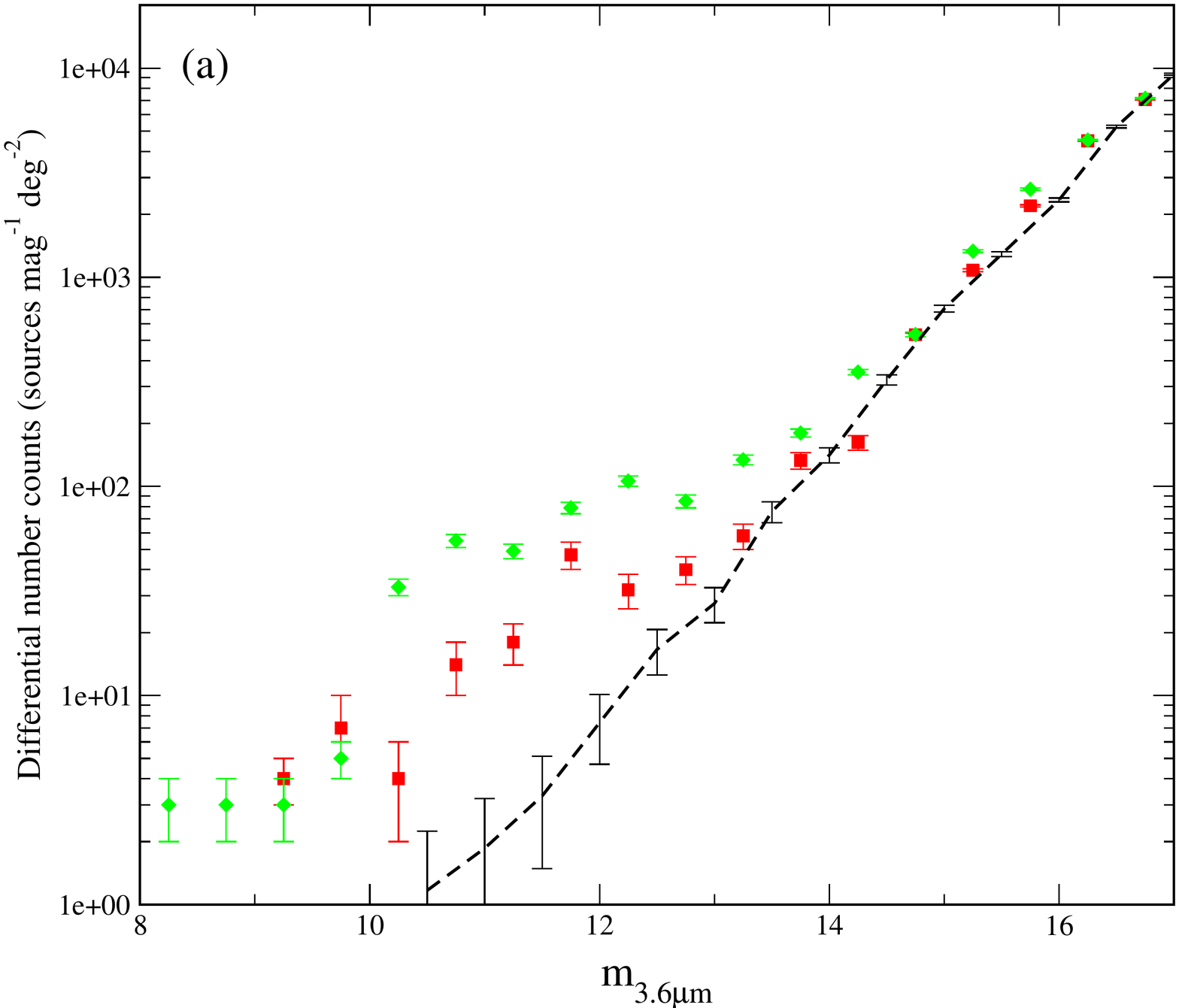}}}
\scalebox{0.35}{\includegraphics{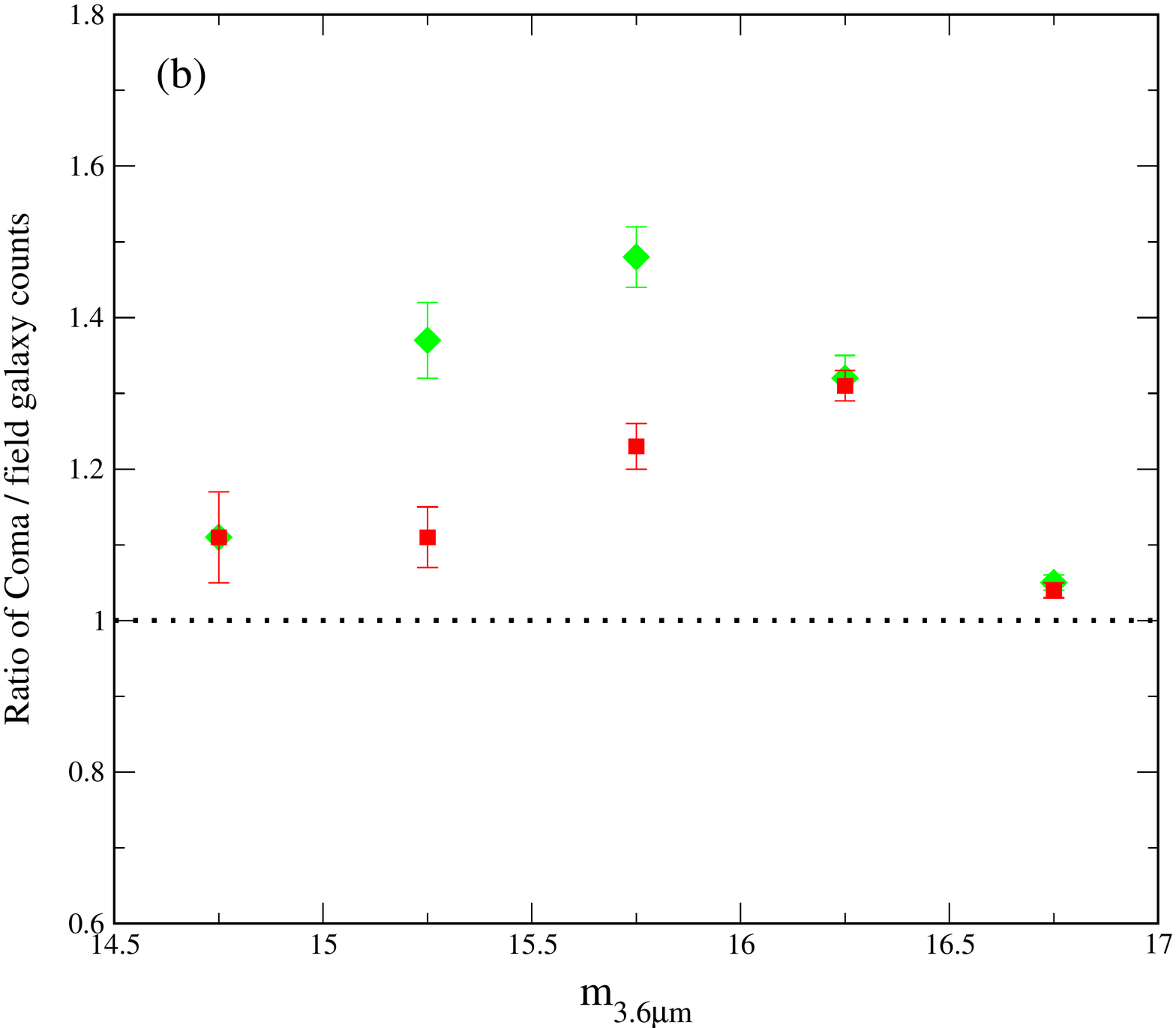}}
\caption{(a) Differential number counts of the IRAC 3.6$\mu$m sources classed as galaxies in the Coma 1 core field ({\it green diamonds}) and the Coma 3 off-center field ({\it red squares}).  Sources classified as stars or cosmic rays have been removed using the stellar concentration parameter and the DIRBE Faint Source Model.   The dashed line shows the field galaxy counts of \cite{fazio04b}; a clear excess of galaxies is evident in both Coma fields with respect to the field population at bright magnitudes. (b) The ratio of galaxies detected in the Coma data to the field galaxy counts of \cite{fazio04b} for the faint magnitude range $14.5<$ m$_{3.6\mu m}<17$, demonstrating that there is also an excess of faint galaxies in Coma with respect to the field.}
\label{fig:galcounts}
\end{center}
\end{figure}

\subsection{Comparison with field galaxy counts}
\label{sec:fieldcounts}

In Figure~\ref{fig:galcounts} (a) we plot the Coma galaxy number counts (excluding all contributions from stars) together with the field galaxy number counts of \cite{fazio04b}.  These consist of observations from 3 fields: QSO 1700 ($5\arcmin\times10\arcmin$), Extended Groth Strip (EGS) ($0\fdg17\times2\degr$) and Bo\"{o}tes ($3\degr\times3\degr$). For magnitudes brighter than 15 mag, we use the wide-area Bo\"{o}tes field counts; fainter than this we use the average of the counts from the deeper QSO 1700 and EGS survey fields (taken from table 1 in \citealt{fazio04b}).  This figure shows that there is a clear excess of galaxies in both Coma fields at bright magnitudes (m$_{3.6\mu m}  <$ 14.5) as expected , but it also indicates an excess at fainter magnitudes.  To illustrate this more clearly, Figure~\ref{fig:galcounts} (b) shows the ratio of galaxy counts in the two Coma fields to the field counts in the range 14.5 $<$ m$_{3.6\mu m} <$ 17.   Even taking possible cosmic field-to-field variations into account (estimated as 16-19\% over 1100 arcmin$^2$ by \citealt{dahlen05}, similar to the EGS deep survey area of 1224 arcmin$^2$), the excess is still above the highest background count estimate.  Note, however, that since we know that there is significant background structure behind Coma, we do not use statistical background source subtraction for the LFs in this study.  Instead, we use spectroscopic membership fractions together with optical color selection to subtract the background galaxy component (see \S~\ref{sec:members} \& \S~\ref{sec:colors}).

\begin{figure}
\begin{center}
\scalebox{0.35}{\includegraphics{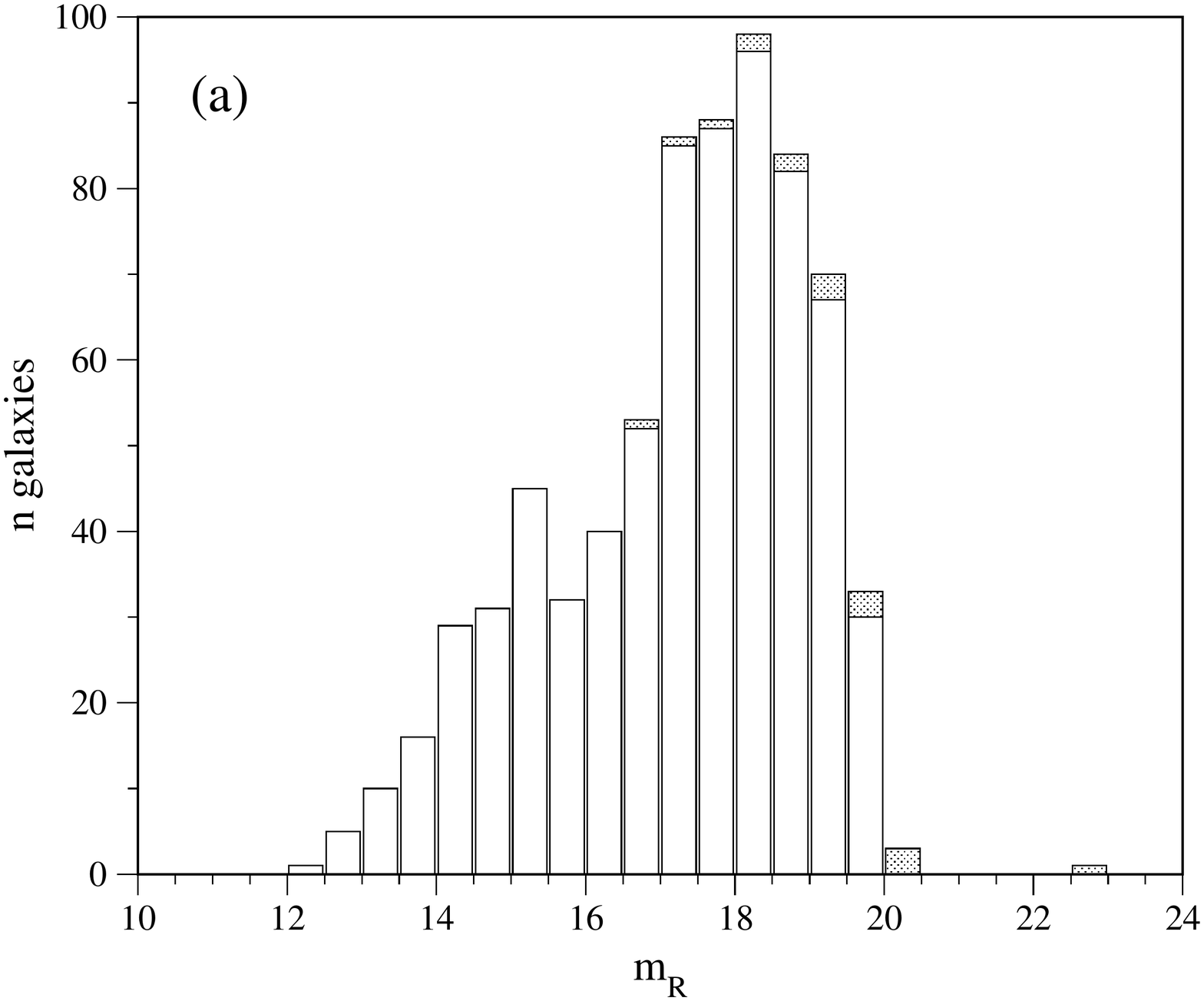}}
\scalebox{0.35}{\includegraphics{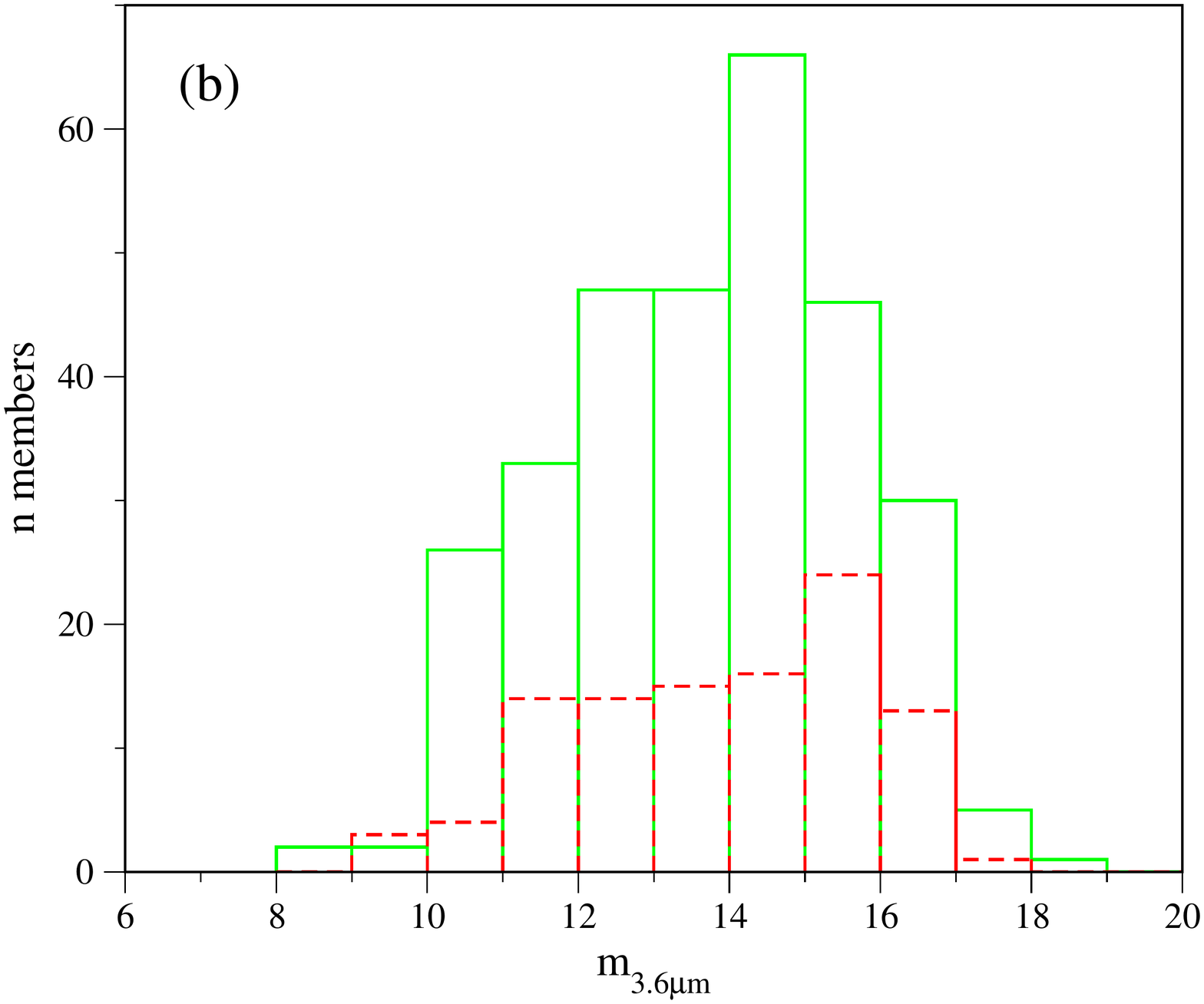}}
\caption{(a) The $R$-band magnitude distribution of the optical spectroscopic targets in the combined \cite{mobasher01} and \cite{colless96} catalogs for the two Coma fields.   The total number of target galaxies per 0.5 mag bin is shown with a shaded histogram, while the number of IRAC detected galaxies matched to the $R$-band sample is overlaid with an open histogram.  Almost all (97\%) of the $R$-band sample is detected at 3.6$\mu$m.  (b) The 3.6$\mu$m magnitude distribution of the spectroscopically confirmed members in the central Coma 1 data ({\it solid line}) and the Coma 3 off-center data ({\it dashed line}).}
\label{fig:stats}
\end{center}
\end{figure}

\section{Optical Data}
\label{sec:optical_survey}

To determine the contribution to the overall LF from background galaxies, we use redshift information from two optical catalogs.  The first is the spectroscopic survey of \cite{mobasher01}, based on the William Herschel Telescope (WHT) wide-field photometric $B$- and $R$-band CCD survey of five fields in the Coma cluster by  \cite{komiyama02}.  In brief, two of the photometric survey fields, each 32\farcm5$\times$50\farcm8, were chosen for follow-up spectroscopic observations using the WYFFOS multi-fiber spectrograph on the WHT; Coma 1 (core), centered on 12$^{h}$59$^{m}$23\fs7, +28\degr01\arcmin12\farcs5, and Coma 3 (off-center), centered on 12$^{h}$57$^{m}$07\fs5, +27\degr11\arcmin13\farcs0.  In this survey, redshifts were measured for two classes of galaxies in Coma: the ``bright'' sample, consisting of galaxies already confirmed as spectroscopic members with previously measured redshifts, and a ``faint'' sample of galaxies with no known redshifts, which were on average fainter than the bright sample. The faint sample was chosen from the photometric data using their positions in the color-magnitude plane ($B-R)$ versus $R$ with the criterion 17.5 $< R <$ 20 (designed to include the dwarf galaxy population) and 1 $< B-R <$ 2 (to eliminate the majority of background galaxies).  

We supplement this with a wider-field (2.63 deg$^2$) but shallower ($R \sim 18$) catalog compiled by M. Colless ({\it private communication}), to ensure the inclusion of previously-studied brighter galaxies in the cluster that were not re-measured in the \cite{mobasher01} study.  This supplemental catalog includes redshift measurements by \cite{colless96} using the Hydra multi-fiber spectrograph at KPNO, measurements from the Two-Degree Field (2dF) survey of \cite{edwards02}, and supplemental redshifts from the NASA/IPAC Extragalactic Database (NED) and literature (\citealt{kent82}, \citealt{caldwell93}, \citealt{haarlem93}, \citealt{biviano95}).  In total, these combined catalogs provide spectroscopic redshift measurements of 727 galaxies across the Coma 1 and Coma 3 fields.

Note that due to the color selection of the spectroscopic sample, we also use deeper photometric data ($r\sim22.5$ mag) from the recent Sloan Digital Sky Survey (SDSS) data release 5 (DR5) to impose a similar optical color cut (SDSS $g-r<1.3$) on our IRAC sample when constructing the 3.6$\mu$m faint-end LFs (see \S~\ref{sec:colors}).

\section{IRAC Cluster Membership \& Spectroscopic Membership Fractions}
\label{sec:members}

\begin{deluxetable}{clccccccc}
\tablecaption{Cluster membership fractions for Coma 1, Coma 3 and the total of the two
\label{tab:members}}
\tablecolumns{9}
\tabletypesize{\small}
\tablewidth{0pt}
\tablehead{
\colhead{} & \multicolumn{2}{c}{Coma 1 core} & \colhead{} &  \multicolumn{2}{c}{Coma 3 off-center} & \colhead{} &  \multicolumn{2}{c}{Total}
\\
\cline{2-3}\cline{5-6}\cline{8-9}
\colhead{m$_{3.6\mu m}$} & \colhead{$f$(m)} & \colhead{$\Delta$ [$f$(m)]} & \colhead{} & \colhead{$f$(m)} & \colhead{$\Delta$ [$f$(m)]}  & \colhead{} & \colhead{$f$(m)} & \colhead{$\Delta$ [$f$(m)]} }
\startdata
9.00........     &  1.00  &  0.00  &  &  1.00  &  0.00 &  &  1.00  &  0.00  \\
10.25......     &  1.00  &  0.00  &  &  1.00  &  0.00 &  &  1.00  &  0.00  \\
10.75......     &  1.00  &  0.00  &  &  1.00  &  0.00 &  &  1.00  &  0.00  \\
11.25......     &  1.00  &  0.00  &  &  0.80  &  0.18 &  &  0.95  &  0.05  \\
11.75......     &  1.00  &  0.00  &  &  0.91  &  0.09  &  &  0.97  &  0.03 \\
12.25......     &  1.00  &  0.00  &  &  1.00  &  0.00  &  &  1.00  &  0.00 \\
12.75......     &  0.81  &  0.09  &  &  0.60   & 0.15   &  &  0.74  &  0.08 \\
13.25......     &  0.79  &  0.08  &  &  0.75   & 0.15   &  &  0.78  &  0.07 \\
13.75......     &  0.73  &  0.07  &  &  0.50  &  0.12  &  &   0.65  &  0.06  \\
14.25......     &  0.49  &  0.06   &  &  0.32  &  0.11  &  &  0.45  &  0.05  \\
14.75......     &  0.59  &  0.07   &  &  0.37  &  0.09  &  &  0.52  &  0.05  \\
15.25......     &  0.44  &  0.07   &  &  0.37  &  0.08  &  &  0.41  &  0.05  \\
15.75......     &  0.40  &  0.07   &  &  0.39  &  0.09  &  &  0.40  &  0.05  \\
16.25......     &  0.50  &  0.09   &  &  0.42  &  0.11   &  &  0.47  &  0.07 \\
16.75......     &  0.57  &  0.09   &  &  0.25  &  0.10   &  &  0.44  &  0.07 \\
17.25......     &  0.27  &  0.11   &  &  0.20  &  0.18   &  &  0.25  &  0.10 \\
17.75......     &  0.33  &  0.27   &  &  \nodata & \nodata  &  &  \nodata & \nodata  \\
\enddata
\end{deluxetable}

In order to identify cluster membership, we cross-correlated the IRAC catalogs with the two optical Coma spectroscopic catalogs discussed in \S~\ref{sec:optical_survey}.  The positions of both the IRAC and optical catalogs are good to within $\la$~1\arcsec, and a 3$\sigma$ search radius of 4.24\arcsec\ was used for the positional matching, corresponding to three times the two positional errors added in quadrature.  Figure~\ref{fig:stats} (a) shows the $R$-band magnitude distribution of the spectroscopic targets in the Coma 1 core and Coma 3 off-center fields, plus the numbers of those galaxies detected in the IRAC 3.6$\mu$m data.  The detection rates with IRAC are high; out of a total of 498 spectroscopic target galaxies in the Coma 1 core field, 487 (98\%) were detected by IRAC, while in the Coma 3 off-center field, 221 (97\%) out of 229 were detected.  

In the spectroscopic catalogs, galaxies with recession velocities in the range 4000 km s$^{-1} < cz <$ 10,000 km s$^{-1} $ are considered to be members of Coma, corresponding to the 3$\sigma$ range measured by \citet{colless96}.  In the Coma 1 field, 312 of the galaxies are members, 306 (98\%) of which were detected with IRAC.   In Coma 3, 108 galaxies are members, of which 104 (96\%) had IRAC matches.  The positions of these 410 confirmed members are plotted on the IRAC 3.6\,$\mu$m mosaic in Figure~\ref{fig:irac_mosaic}, and their 3.6$\mu$m magnitude distributions are shown in Figure~\ref{fig:stats} (b).  IRAC finds almost all of the known optical members.  Note that there is some weak IR emission at the locations of the few undetected members in the 3.6$\mu$m images (all fainter than $R=16.5$), but not significant enough to be detected above the 3$\sigma$ threshold.

\begin{figure}
\begin{center}
\scalebox{0.43}{\includegraphics{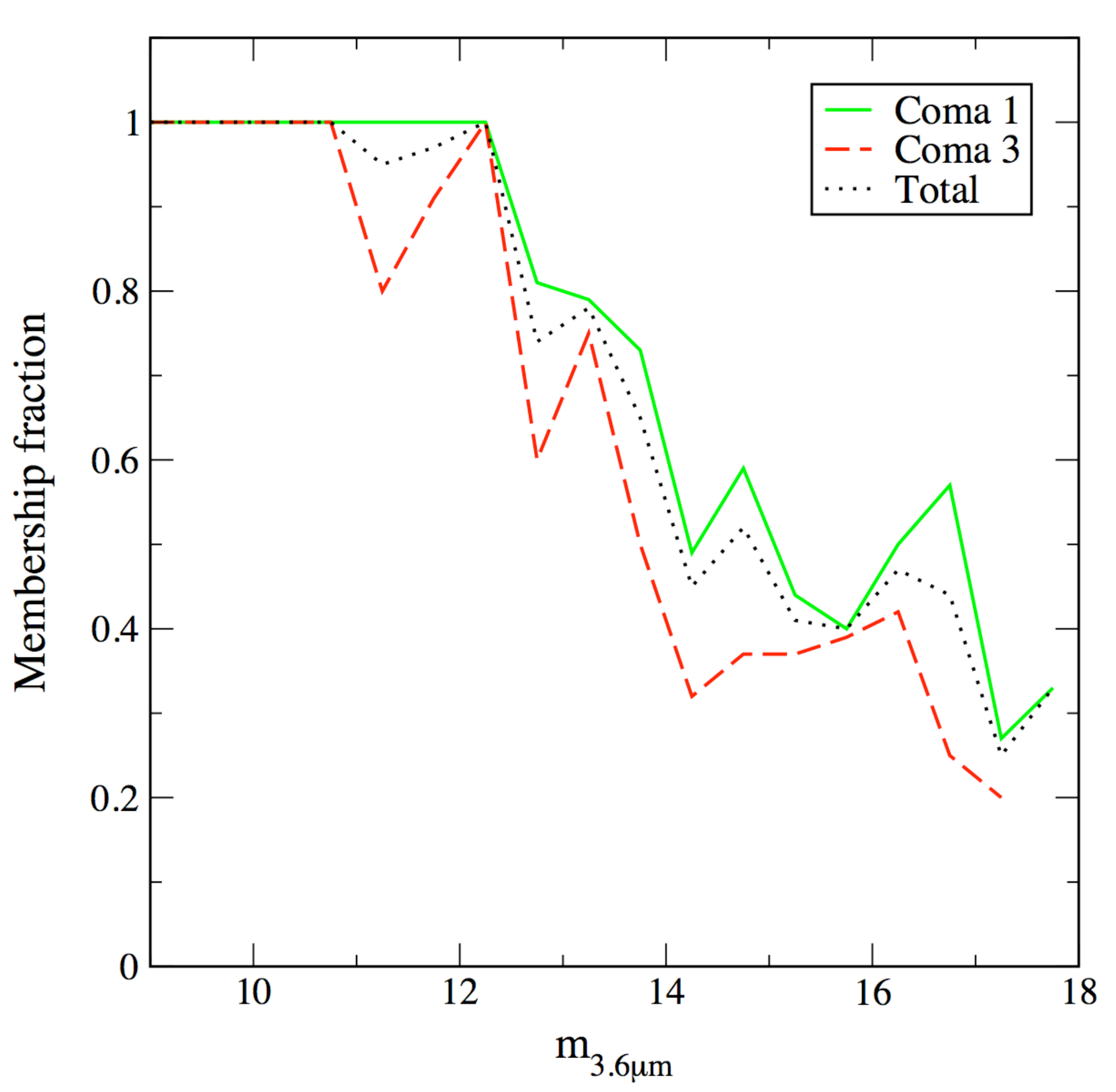}}
\caption{The spectroscopic membership fractions of the IRAC 3.6\,$\mu$m detected galaxies as a function of apparent magnitude for the Coma 1 and Coma 3 fields, and the total of the two.}
\label{fig:complete}
\end{center}
\end{figure}

\begin{deluxetable*}{lcccccccc}
\tablecaption{Luminosity functions for Coma 1, Coma 3 and the total of the two 
\label{tab:lf}}
\tablecolumns{9}
\tablewidth{0pt}
\tablehead{
\colhead{} & \multicolumn{2}{c}{Coma 1 (core)} & \colhead{} &  \multicolumn{2}{c}{Coma 3 (off-center)} & \colhead{} & \multicolumn{2}{c}{Total}
\\
\cline{2-3}\cline{5-6}\cline{8-9}
\colhead{M$_{3.6\mu m}$} & \colhead{$\phi$(M)} & \colhead{$\Delta$ [$\phi$(M)]} & \colhead{} & \colhead{$\phi$(M)} & \colhead{$\Delta$ [$\phi$(M)]} &\colhead{} & \colhead{$\phi$(M)} & \colhead{$\Delta$ [$\phi$(M)]}  }
\startdata

\cutinhead{All IRAC detections (no optical constraint)}

-26.00..........     &  0.56      &  0.50     &  &   0.44     &  0.51      &  &    0.51      &  0.36   \\
-24.75..........     &  5.37      &  1.55     &  &   0.59     &  0.59      &  &    3.31      &  0.92   \\ 
-24.25..........     &  8.95      &  2.00     &   &  2.36     &  1.18      &  &    6.11      &  1.25   \\
-23.75..........     &  8.06      &  1.90     &   &  2.36     &  1.18      &  &    5.55      &  1.20   \\
-23.25..........     &  12.98    &  2.41     &   &  6.98     &  2.05      &  &   10.34    &  1.63   \\
-22.75..........     &  17.46    &  2.80     &  &   5.32     &  1.77      &  &   12.22    &   1.76  \\
-22.25..........     &  11.23    &  2.34     &  &   3.90     &  1.55      &  &    7.94      &   1.49    \\
-21.75..........     &  17.36    &  3.08     &  &   7.09     &  2.29      &  &    12.93    &   2.01   \\
-21.25..........     &  20.58    &  3.31     &  &   10.63   &  3.07      &  &    16.50    &   2.32    \\
-20.75..........     &  28.48    &  4.30     &  &   8.02     &  2.97      &  &    19.91    &   2.78    \\
-20.25..........     &  50.65    &  6.73     &  &   31.28   &  8.27      &  &    44.20    &   5.27    \\
-19.75..........     &  94.84    &  15.39   &  &   64.06   &  14.58    &  &    81.24   &   10.77    \\
-19.25..........     &  169.38  &  28.51   &  &   140.85 &  33.58    &  &   157.23  &  21.62    \\
-18.75..........     &  368.62  &  67.91   &  &   308.89 &  83.56    &  &   345.32  &  52.84     \\
-18.25..........     &  670.16  &  110.46 &  &   \nodata &  \nodata  &  &   \nodata  &  \nodata     \\

\cutinhead{$g-r<2$ ($r<22.5$) faint end}

-20.25..........     &  45.63    &  6.16     &  &   29.75   &  7.89      &  &    40.77    &   4.90    \\
-19.75..........     &  75.63    &  12.40   &  &   51.56   &  11.83    &  &    65.01   &   8.70    \\
-19.25..........     &  111.73  &  18.99   &  &   100.01 &  23.99    &  &   106.81  &  14.80    \\
-18.75..........     &  185.54  &  34.48   &  &   155.44 &  42.27    &  &   173.79  &  26.79     \\
-18.25..........     &  235.84  &  39.37 &  &   \nodata &  \nodata  &  &   \nodata  &  \nodata     \\

\cutinhead{$g-r<1.3$ ($r<22.5$) faint end}

-20.25..........     &  19.78    &  3.18     &  &   11.16   &  3.21        &  &    16.62    &   2.30    \\
-19.75..........     &  34.85    &  6.03     &  &   21.72   &  5.25        &  &     28.98   &   4.09    \\
-19.25..........     &  63.92    &  11.09   &  &   50.82   &  12.42      &  &    58.32   &  8.25    \\
-18.75..........     &  117.05  &  21.98   &  &   95.75   &  26.22      &  &    108.53  &  16.87     \\
-18.25..........     &  160.89  &  27.10   &  &   \nodata &  \nodata  &  &   \nodata  &  \nodata     \\

\enddata

\tablecomments{~LFs are in units of galaxies 0.5 mag$^{-1}$ Mpc$^{-2}$.  A distance modulus of 35.0 mag ($H_0$=70\,km s$^{-1}$ Mpc$^{-1}$) is assumed for Coma.}

\end{deluxetable*}

To construct LFs for the Coma galaxies, we must estimate the contribution of background galaxies to the IRAC number counts.  To achieve this, we compute a cluster membership fraction for each field. The optical spectroscopic coverage of Coma is not complete, but the \citet{mobasher03} spectroscopic survey was specifically designed to sample optical color magnitude space in a way that determines membership within each half-magnitude range down to $R\sim20$.  Thus, although not every galaxy down to $R=20$ has a spectrum, the membership of the cluster is constrained to this depth.  Therefore, we assume that the incomplete spectroscopic sample is representative of the whole galaxy population in the Coma cluster, so that the fraction of galaxies determined to be cluster members in the spectroscopic survey as a function of magnitude can be applied to the complete catalog of IRAC detected galaxies.  The magnitude limits of the IRAC and optical spectroscopic catalogs are well matched for determining this function: the faintest 3.6$\mu$m bins correspond to m$_{R}\sim20-21$, the limit of the spectroscopic data.

We define the spectroscopic membership function ({\it f}, Eqn.~\ref{equ:f}) as the ratio of the number of spectroscopically confirmed cluster members in a given apparent magnitude bin, $N_{c}$(m$\mid$spec),  to the total number of galaxies in the spectroscopic catalog (members and non-members), detected by IRAC, in the same magnitude bin, N$_{t}$(m$\mid$spec):

\begin{equation}
f (m\mid spec) = \frac{N_{c}(m\mid spec)}{N_{t}(m\mid spec)}
\label{equ:f}
\end{equation}

This ratio is calculated separately for the Coma 1 core and Coma 3 off-center fields (and total of the two).  The relative error on these functions, assuming poisson statistics, is given by:

\begin{equation}
\frac{df (m\mid spec)}{f (m\mid spec)} = \left[\frac{1}{N_{c}(m\mid spec)} - \frac{1}{N_{t}(m\mid spec)}\right]^{\frac{1}{2}}
\end{equation}

The membership fractions were determined in half-magnitude bins for galaxies with m$_{3.6\mu m} > 10$.  At the bright end, where there are far fewer galaxies, they were binned between 8 $<$ m$_{3.6\mu m}   < 10$ to ensure that the statistics were good enough for constraining the data.  These membership fractions (and errors) per bin for the two fields and for the total of the two are shown in Table~\ref{tab:members} and plotted in Figure~\ref{fig:complete}.  The Coma 1 data extends down to m$_{3.6\mu m}=18$, while the Coma 3 is limited to m$_{3.6\mu m}=17.5$ due to the slightly shallower observation, although we impose slightly brighter magnitude limits for the LF calculations (see \S~\ref{sec:lf}).  As expected, the fraction of member galaxies per magnitude bin is higher in the Coma 1 field due to the relatively larger number of brighter galaxies at the center of the cluster.

\section{The Luminosity Functions}
\label{sec:lf}

In this section we present the IRAC 3.6$\mu$m LF of galaxies in the Coma cluster.  The LFs for the Coma 1 and Coma 3 fields are constructed separately, using the membership fractions found in the last section.  These are estimated as:

\begin{equation}
\phi (m) = \frac{N_{p}(m)}{A}    f (m\mid spec)
\label{equ:phi}
\end{equation}

\noindent where $N_{p}(m)$ are the galaxy counts, $f$ is the membership fraction and $A$ is the area of the cluster over which the counts are measured.  To improve statistics on the faint end of the IRAC LF, we extend our sample beyond the regions for which spectroscopic data are available.  The relative errors on the LFs are given by:

\begin{equation}
\frac{d\phi (m)}{\phi(m)} = \left[\frac{1}{N_{p}(m)} + \frac{1}{N_{c}(m\mid spec)} - \frac{1}{N_{t}(m\mid spec)} \right]^{\frac{1}{2}}
\end{equation}

\begin{figure*}
\begin{center}
\scalebox{0.7}{\includegraphics{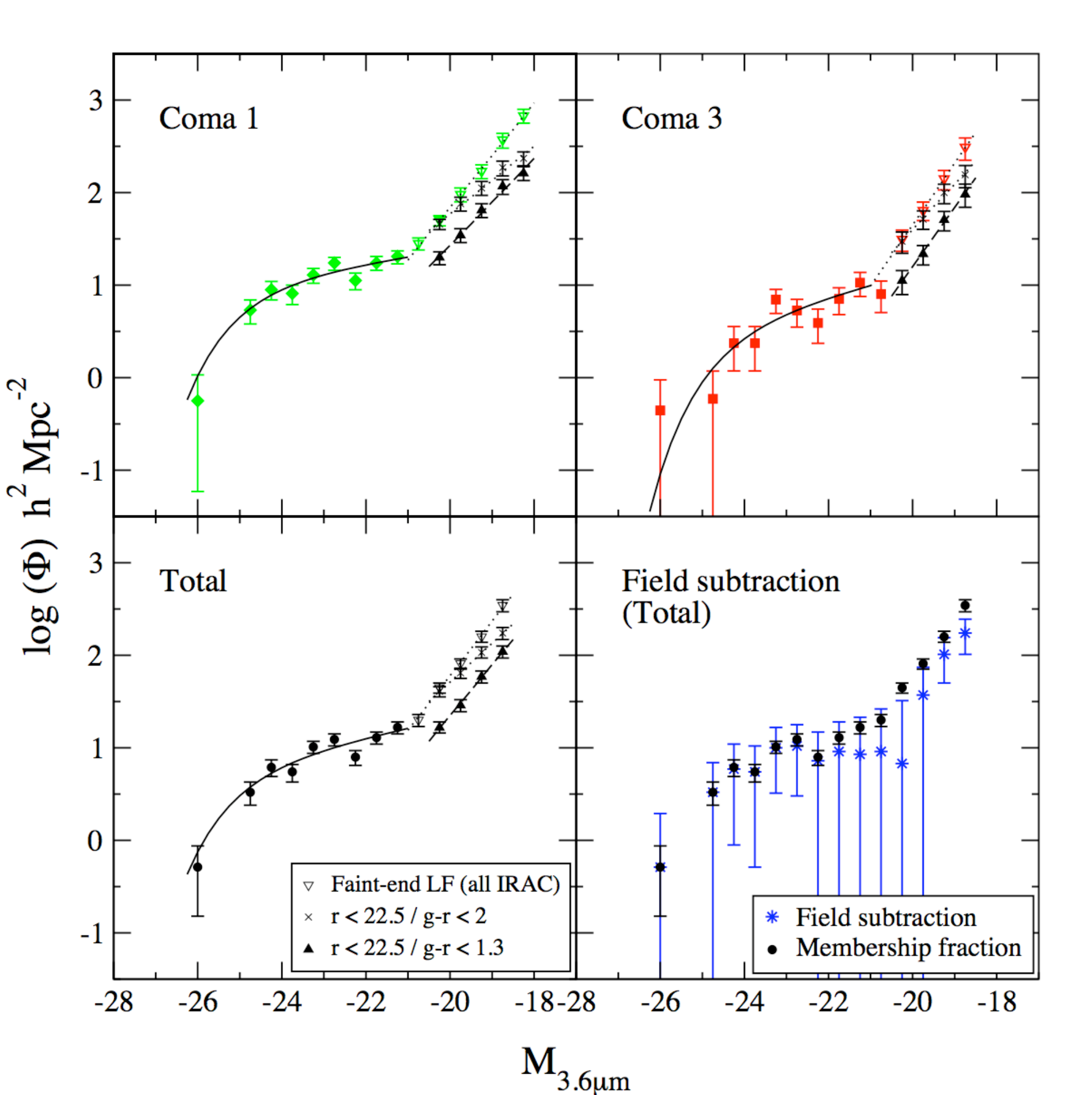}}
\caption{The 3.6$\mu$m LFs for the Coma 1 core field (top left, {\it green}), the Coma 3 off-center field (top right, {\it red}) and the total of the two (lower left, {\it black}).  The best-fit Schechter functions, listed in Table~\ref{tab:fits}, are shown at the bright end of each LF (m$_{3.6\mu m} < 14.5$).  At the faint end (m$_{3.6\mu m} > 14.5$), we show the `raw' LFs, which take into account all IRAC-detected galaxies, and which we consider to be upper limits to the faint-end normalization.  We also show the two $g-r$ color-selected faint-end LFs for the IRAC sources with SDSS optical counterparts down to $r=22.5$, with the $g-r<1.3$ selection being the most conservative.  The faint-end power-law fits, also listed in Table~\ref{tab:fits}, are shown with straight lines.  Note that although the overall number of galaxies decreases with these optical selections, the faint-end upturn is still present.  In the lower right panel, we also show a comparison between different methods of background subtraction for the total LF (membership fraction and field subtraction); the two are consistent in almost all bins, and the faint-end upturn is present in the field-subtracted LF.}
\label{fig:lf_all}
\end{center}
\end{figure*}

At a distance of 100\,Mpc, 1\degr = 1.75\,Mpc, and the areas covered by the IRAC observations correspond to 2.234\,Mpc$^2$ for the Coma 1 core field, and 1.693\,Mpc$^2$ for the Coma 3 off-center field.  ``Raw" LFs for both fields, calculated using all IRAC-detected galaxies, are listed in Table~\ref{tab:lf} and plotted in Figure~\ref{fig:lf_all}.  However, in the next section we impose optical color selections to the faint sample to create optically-constrained faint-end LFs, and we therefore consider the faint end of these raw LFs  (M$_{3.6\mu m}>-20.5$) to be upper limits only.  The Coma 1 LF covers the absolute magnitude range of $-27 <$ M$_{3.6\mu m} < -18$ (using a distance modulus of 35.0), while for Coma 3 the LF extends down to M$_{3.6\mu m} = -18.5$.  Within these magnitude ranges, these initial calculations imply a total of 3320 IRAC-selected members in the Coma 1 core field and 1006 in the Coma 3 off-center field.  This includes both the known spectroscopic members and additional galaxies with no redshift information (see Table~\ref{tab:fits}).  The LFs for both fields are consistent with a Schechter form, with a striking upturn at M$_{3.6\mu m}\sim$-20.5, presumably caused by the onset of the dwarf galaxy population.

\begin{figure}
\begin{center}
\scalebox{0.45}{\includegraphics{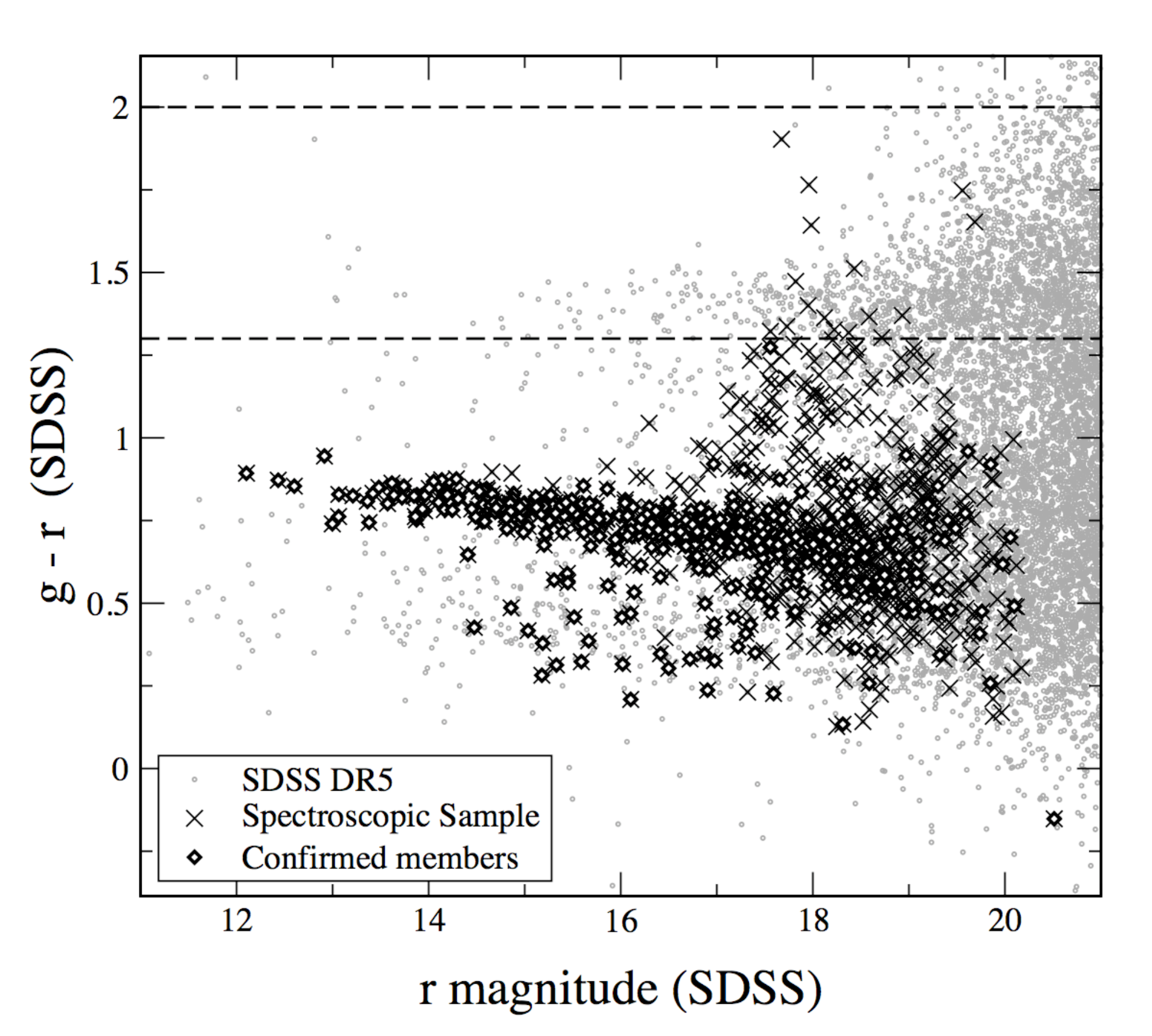}}
\caption{The $g-r$ versus $r$ color-magnitude relation for the IRAC 3.6$\mu$m detected galaxies matched to optical sources with $r<22.5$ in the SDSS DR5 catalog.  The Coma member galaxies form a strong red sequence at $0.5 < g-r < 1$, and also note the red sequence of the background cluster at $g-r\sim1.4$.  The dashed horizontal lines denote the two optical color cuts applied to the faint end of the LF.  The $g-r<2$ cut is used to correct for any optical bias in the spectroscopic completeness function, and corresponds to the color space sampled by the \cite{mobasher01} spectroscopic survey.  The $g-r<1.3$ cut is our most conservative, and is chosen to subtract the contribution of the background cluster and to be consistent with the SDSS color-magnitude distribution of galaxies in the local Universe \citep{hogg04}.}
\label{fig:sdss}
\end{center}
\end{figure}

\begin{deluxetable*}{lccccc}
\tablecaption{Schechter and power-law fits to the Coma 1, Coma 3 and total LFs
\label{tab:fits}}
\tablecolumns{6}
\tabletypesize{\small}
\tablewidth{0pt}
\tablehead{
\multicolumn{1}{l}{Optical color selection} & \colhead{M$^{\star}_{3.6\mu m}$} &  \colhead{Schechter slope ($\alpha_1$)} & \colhead{Faint-end slope ($\alpha_2$)} &  \colhead{$N_{gal}$}  &  \colhead{$N_{member}$ (inferred)}
\\
\multicolumn{1}{l}{(1)} & \colhead{(2)} & \colhead{(3)} & \colhead{(4)} & \colhead{(5)} & \colhead{(6)} }

\startdata

\cutinhead{{\bf Coma 1 (core):}\hspace{0.2cm}$-27 < $M$_{3.6\mu m} < -18$}

None                                   &  -25.17$\pm$0.10  &  -1.18$\pm$0.06  &  -2.40$\pm$0.13  &  6280  &  3320  \\
$g-r<2$ ($r<22.5$)     &  \nodata                    &  \nodata                 &  -1.93$\pm$0.13  &  3326  &  1757  \\
$g-r<1.3$ ($r<22.5$)  &  \nodata                    &  \nodata                 &  -2.18$\pm$0.13  &  2156  &  1181  \\

\cutinhead{{\bf Coma 3 (off-center):}$\hspace{0.2cm}-27 < $M$_{3.6\mu m} < -18.5$}

None				  &  -24.69$\pm$0.15  &  -1.30$\pm$0.08  &  -2.68$\pm$0.30  &  2425  &  1006  \\
$g-r<2$ ($r<22.5$)     &  \nodata                    &  \nodata                 &  -2.23$\pm$0.30  &  1568  &  653  \\
$g-r<1.3$ ($r<22.5$)  &  \nodata                    &  \nodata                 &  -2.60$\pm$0.33  &  895  &  387  \\

\cutinhead{{\bf Total:}\hspace{0.2cm}$-27 < $M$_{3.6\mu m} < -18.5$}

None				   &  -25.28$\pm$0.20   &  -1.25$\pm$0.05  &  -2.48$\pm$0.15  &  6086  &  2846  \\
$g-r<2$ ($r<22.5$)      &  \nodata                    &  \nodata                 &  -2.05$\pm$0.15  &  3973  &  1898  \\
$g-r<1.3$ ($r<22.5$)   &  \nodata                    &  \nodata                 &  -2.38$\pm$0.15  &  2423  &  1215  \\

\enddata

\tablecomments{(1) SDSS optical color selection of the faint (M$_{3.6\mu m}> -20.5$) IRAC 3.6$\mu$m galaxies. ``None" refers to IRAC LFs with no optical constraint.  (2)  Bright magnitude turnover in the Schechter function.  (3)  Schechter slope, $\alpha_1$, of bright end of the LF (M$_{3.6\mu m} < -20.5$).  (4) Faint-end slope (M$_{3.6\mu m} > -20.5$), where $\alpha_2=-(k/0.4 + 1)$ and $k$ is the logarithmic power-law slope.  (5)  Total numbers of galaxies detected at 3.6$\mu$m over given magnitude ranges, excluding all stars (morphological/statistical).  (6)  Total numbers of {\it member} galaxies inferred by the spectroscopic membership fractions and optical color-cuts over the given magnitude ranges.}

\end{deluxetable*}

\subsection{Optical constraints}
\label{sec:colors}

As mentioned in \S~\ref{sec:optical_survey}, the spectroscopic sample used to define the membership fraction is selected based on optical color criteria. Therefore, we have used optical data from the SDSS DR5, which covers both Coma fields, to create two subsets of IRAC-detected sources filtered by their optical properties ($g-r$ color and $r$-band magnitude) that are more closely matched to the optically-selected population on which spectroscopy was performed.   We seek to strike a balance between guarding against this optical selection bias and losing valuable information afforded by our IR selection.  We first require that the IRAC sources have optical counterparts above the SDSS photometric limit of $r\sim22.5$ (i.e., we do not include any optically {\it undetected} sources in our LFs).    Inspection of the $g-r$ versus $r$ plot in Figure~\ref{fig:sdss} shows that the optical spectroscopy of Mobasher et al. (2001) samples this color-magnitude space to $g-r\sim2$.   Note that there is a known background cluster behind Coma at $z\sim0.5$ \citep{adami00}, whose red sequence  is clearly visible in Figure~\ref{fig:sdss} around $g-r\sim1.4$.  Using the color-magnitude distribution of galaxies in the local Universe found in \cite{hogg04}, we estimate $\sim97$\% of galaxies to have $g-r<1.2$.   With all these considerations in mind, we find that $g-r<1.3$ is the best compromise color, which we term ``conservative",  in that it is the most strict color filter that does not {\it over-subtract} the galaxies but also guards against contamination by the background cluster.    We have also included an ``optimistic" $g-r<2$ sample, corresponding to the color space sampled spectroscopically by \cite{mobasher01}.  

Overall, $\sim$60\% of IRAC sources above the adopted 3.6$\mu$m LF completeness limits have optical counterparts down to $r=22.5$.  When the color criteria are applied, there are 4,894 3.6$\mu$m galaxies in the optimistic sample, and 3,051 galaxies in the conservative sample, after subtraction of the faint-end stellar contribution (see Table~\ref{tab:fits}).  We choose to only apply the color cuts to the faint end of the LFs (m$_{3.6\mu m}>14.5$), where the largest uncertainty in the cluster membership fraction exists (due to poorer spectroscopic sampling statistics).  When the spectroscopic membership fractions are applied to the $g-r<2$ filtered galaxy counts,  a significantly reduced number of members is implied: 1757 in Coma 1 ($-27 <$ M$_{3.6\mu m} < -18$) and 653 in Coma 3 ($-27 <$ M$_{3.6\mu m} < -18.5$). For $g-r<1.3$, the member totals are further reduced to 1181 in Coma 1 and 387 in Coma 3.  The two sets of optically filtered LFs are plotted for both fields in Figure~\ref{fig:lf_all}, together with the optically unconstrained ``raw" IRAC LFs to demonstrate the plausible ranges of slope and normalization.  {\it Note that although the numbers of member galaxies are reduced by the color-cuts, the rise at the faint end is still significant in both fields.}

\subsection{Parametric Fits}

The widely used parametric form to describe the shape of the galaxy LF is the Schechter function \citep{schechter76}:

\begin{equation}
\Phi (M) =   \phi^{\star} X^{\alpha +1} e^{-X}
\label{equ:schechter}
\end{equation}

\noindent where $X = 10^{-0.4(M-M^{\star})}$. $M^{\star}$ is the characteristic magnitude where the distribution turns over at the bright end due to the sharp decline in the numbers of galaxies at these magnitudes, $\alpha$ is the logarithmic faint-end slope, and $\phi^{\star}$ is the normalization. 

A lot of studies of the Coma LF fit a single Schechter function to the data.   However, the sharp rise at the faint end of the 3.6$\mu$m LF in both fields makes it obvious that a single function will not be adequate in this case.  Instead, we choose to fit a Schechter function to the bright end (M$_{3.6\mu m}<-20.5$) to model the $M^{\star}$ and $\alpha$ parameters, and an exponential power-law function with the following form to model the dwarf galaxy population at the faint end (M$_{3.6\mu m}>-20.5$):

\begin{equation}
\Phi (M) \propto 10^{kM}
\label{equ:plaw}
\end{equation}

\noindent where $k$ is the power-law slope, which can be transformed into the equivalent of the Schechter function slope via $\alpha=-(k/0.4+1)$ (e.g. \citealt{sandage85}).  To distinguish between the slopes of the two components, we refer the Schechter slope as $\alpha_1$ and the faint-end power-law slope as $\alpha_2$. 

The results of the fits are shown in Table~\ref{tab:fits} for the two Coma fields (and their total), and are plotted in Figure~\ref{fig:lf_all}.  At the bright end, the best-fitting Schechter slopes range between fairly flat values of $\alpha_1$ = -1.18$\pm$0.06 (Coma 1) and -1.30$\pm$0.08 (Coma 3), with M$^{\star}_{3.6\mu m}$ values ranging between -25.17$\pm$0.10 and -24.69$\pm$0.15 respectively.  At the faint end, the raw IRAC LFs are very steep with slopes $\alpha_2$ = -2.40$\pm$0.13 (Coma 1) and -2.68$\pm$0.30 (Coma 3).  With the most conservative optical color-cut imposed ($g-r<1.3$), these are reduced to $\alpha_2$ = -2.18$\pm$0.13 in Coma 1 and -2.60$\pm$0.33 in Coma 3.  Note that there is a difference in all these parameters between the Coma 1 core and Coma 3 off-center fields, which is indicative of environmental effects (see \S~\ref{sec:environ}).

\section{Discussion}
\label{sec:disc}

This study provides, for the first time, a measure of the 3.6$\mu$m LF for a cluster at $z\sim0$, and indicates that IRAC observations may provide a very efficient way to search for dwarf galaxies in clusters.  It covers a similar absolute magnitude range as IR LFs for galaxies in more distant clusters, allowing study of the evolution of the mass function in the cluster environment.  

The shape of the IRAC 3.6$\mu$m LF at the bright end (M$_{3.6\mu m}<-20.5$) is consistent with a Schechter form.  The characteristic magnitude (M$_{3.6\mu m}^{\star}$) is brighter (by $\sim$ 0.5 mag) for galaxies at the denser core region compared to the off-center population.  Assuming that the 3.6$\mu$m flux measures the underlying mass of galaxies, we conclude that the concentration of massive galaxies at the core of the Coma cluster is presumably caused by mergers of smaller galaxies as they fall in the gravitational field of Coma.  This is supported by the increase in space density of the less massive galaxies (i.e. the Schechter faint-end slope, $\alpha_1$) towards the outskirts of the cluster.  

The important result from this study is depicted in Figures~\ref{fig:lf_all} and \ref{fig:lf_compare}, where we explore the very faint tail of the 3.6$\mu$m LF ($\alpha_2$).  The steep upturn at M$_{3.6\mu m}\sim -20.5$ is very striking, and is present in both the core and off-center regions of the cluster.  As shown in \S~\ref{sec:colors}, a significant fraction ($\sim$60\%) of IRAC-detected galaxies have optical counterparts in the SDSS data down to $r\sim22.5$, which indicates that this sub-sample are not spurious data artifacts.  Using spectroscopic identification and conservative optical color cuts to remove contamination by background objects, the upturn at the faint end remains.  Furthermore, the background cluster at $z\sim0.5$ is accounted for and removed by the $g-r<1.3$ color selection.  Also, an independent technique for background subtraction, using counts from field samples, gives the same trend (see \S~\ref{sec:background}).  

The results from the conservative $g-r<1.3$ color selected IRAC sample infer that there are at least $\sim1600$ member galaxies over the two fields (see Table~\ref{tab:fits}).  A simple extrapolation of these results also implies that there are at least $\sim$ 5300 member galaxies over the whole area of Coma out to the extreme radius of the Coma 3 field ($\sim87\arcmin$), assuming the same number density of galaxies as the Coma 3 field across the unobserved area.  The inferred stellar masses of the galaxies in the upturn portion of the LF is $\sim5\times10^7-3\times10^8$ M$_{\odot}$ (see \S~\ref{sec:compare_other}).  This is comparable to the stellar mass of, for instance, the SMC ($\sim3\times10^8$\,M$_{\odot}$, \citealt{leroy07}), but less massive than the LMC ($\sim2\times10^9$\,M$_{\odot}$, \citealt{schommer92}).  

However, there are a number of selection effects which might lead to this upturn feature, with the most likely being contribution by background objects or cosmic variance.  Bearing in mind these caveats, and assuming the observed upturn to be real, we now investigate the implications of this result by making a detailed comparison with other multi-wavelength studies in the literature.  Studies of the Coma LF over the past $\sim$25 years are numerous, particularly at optical wavelengths, and we have compiled measurements of the LF slope from the literature in Table~\ref{tab:lfsummary}.  Also listed are the areas covered by each study and the method of background source subtraction used. Where single functions were fitted to the data, the value of $\alpha$ reflects the slope over the entire magnitude range above M$^{\star}$ in that particular band.  In cases where only the faint data were fitted, or where two component models were used (either a second Schechter function or a power-law), the quoted value of $\alpha$ is for the faint component only.

\begin{deluxetable*}{lcccccc}
\tablecaption{Other measured slopes of the Coma LF 
\label{tab:lfsummary}}
\tablecolumns{7}
\tabletypesize{\footnotesize}
\tablewidth{18cm}
\tablehead{
\colhead{} & \colhead{} & \colhead{} & \colhead{Field Size} & \colhead{} & \colhead{Background Field Size} & \colhead{}  \\
\multicolumn{1}{l}{Filter} & \colhead{$\alpha$} & \colhead{Magnitude Range$^a$} & \colhead{(arcmin$^2$)} & \colhead{Background Subtraction Method} & \colhead{(arcmin$^2$)} & \colhead{Reference} }
\startdata

\cutinhead{UV \& Optical}

UV                     &   -2.16$^{+0.10}_{-0.10}$   & 13.0-17.0   & 11300      & Redshift          & \nodata  & 1 \\
UV                     &   -1.65$^{+0.30}_{-0.30}$   & 15.5-18.3 (F)   & 3500        & Redshift           & \nodata   & 2 \\

$U$ (core)        &   -1.32$^{+0.02}_{-0.03}$   &  14.2-21.2  &   452        & Control field    &  980  &  3 \\
$U$ (total)        &   -1.54$^{+0.04}_{-0.03}$   &  14.2-21.2   &   4680      & Control field    &  980  &  3 \\
   
$B$ (core)        &   -1.37$^{+0.02}_{-0.02}$   &  13.2-21.2    &   452         & Control field    &  2000  &  3 \\ 
$B$ (total)        &   -1.32$^{+0.06}_{-0.05}$   &  13.2-21.2     &   18720     & Control field    &  2000  &  3 \\ 

$B$ (North)      &   -1.48$^{+0.03}_{-0.03}$   &  $<$24.75       &   1200      & Membership fraction &  10200  &   4  \\
                           &                                                 &                           &                  & \& Control field          &                &       \\ 
$B$ (South)     &   -1.32$^{+0.03}_{-0.03}$   &  $<$24.75        &   800         & Membership fraction  &  10200  &   4  \\ 
                           &                                                 &                           &                   & \& Control field          &                &       \\ 
$b$                    &   $\sim -1.32$                        &  17.5-20.0 (F) &  14300     & $(b-r)$ color   & \nodata   &  5  \\
$b$                    &   -1.3$^{+0.1}_{-0.1}$          &  17.0-20.0 (F) &  1200       &  Redshift \& $(b-r)$ color  &  \nodata  &  6   \\
$B$                    &   -0.96$^{+0.01}_{-0.02}$   & 13.0-20.0    &  4950       &  Membership fraction  &  \nodata  &   7  \\
$B$                    &  $\sim -1.25$                        &  18.0-22.0  (F) &  720         &  Control field           &  612  &  8   \\     

$V$                    &  $\sim -1.4$                          &  16.0-22.0  (F)    & 1044         &  Control field           &  650  &  8   \\  
$V$  (core, N4874) &  -1.58$^{+0.10}_{-0.10}$  &  16.5-21.0 (F) &  51      &  Control field &  1440  &  9  \\
$V$  (core, N4889) &  -1.51$^{+0.13}_{-0.13}$  &  16.5-21.0 (F)  &  51      &  Control field &  1440  &  9  \\
$V$  (total)       &  -1.81$^{+0.03}_{-0.03}$   &  16.5-21.0 (F) &  1422          &  Control field &  1440  &  9  \\

$V$ (North)      &   -1.72$^{+0.01}_{-0.01}$   &  $<$24.0     &   1200      & Membership fraction &  10200  &   4  \\ 
                           &                                                 &                           &                  & \& Control field          &                &       \\ 
$V$ (South)     &   -1.27$^{+0.03}_{-0.03}$   &  $<$24.0       &   800         & Membership fraction  & 10200   &   4 \\ 
                           &                                                 &                           &                  & \& Control field          &                &       \\ 
$r$ (core)         &   -1.16$^{+0.01}_{-0.02}$   &  12.2-21.2  &   452         & Control field    &  1000  &  3 \\ 
$r$ (total)         &   -1.22$^{+0.03}_{-0.04}$   &  12.2-21.2    &   18720     & Control field    & 1000   &  3 \\ 
$r'$                     &  -1.55$^{+0.08}_{-0.08}$   &  16.0-20.5 (F) &  3600        &  Control field    & 3600   & 10 \\
H$\alpha$         &  -0.60$^{+0.07}_{-0.10}$     &  13.0 $< r'<$ 20.0    &  3600        &  Redshift           & \nodata   & 11 \\    

$R$                    &  $\sim -1.7$                          &  20.0-24.0 (F) &  674          & Control field &  188  &  12  \\
$R$                    &  -1.42$^{+0.05}_{-0.05}$   &  15.5-23.5 (F) &  52            & Control field  & 270   &  13  \\ 
$R$                    &  -1.42$^{+0.12}_{-0.12}$   &  16.5-20.0 (F) &  529          & Control field  &  2645  &  14  \\
$R$                    &  -1.41$^{+0.05}_{-0.05}$   &  15.5-22.5 (F) &  700          &  $(B-R)$ color   & \nodata   &  15  \\
$R$ (core)         &   -1.17$^{+0.03}_{-0.02}$  & 12.0-19.0  & 1650        &  Membership fraction  &  \nodata  &   7     \\
$R$ (off-center)  &   -1.29$^{+0.04}_{-0.03}$  & 12.0-19.0   &  1650        &  Membership fraction  &  \nodata  &   7     \\
$R$ (total)         &   -1.18$^{+0.04}_{-0.02}$   & 12.0-19.0   &  4950        &  Membership fraction  &  \nodata  &   7     \\  
$R$                    &  $\sim -1.4$                            & 16.0-22.0 (F) & 1044       &  Control field           &  650  &  8   \\ 
$R$                    &  -2.29$^{+0.33}_{-0.33}$     &  22.0-26.0 (F) &  5              &  Control field      &  5  &   16 \\  

$R$ (North)      &   -1.74$^{+0.02}_{-0.02}$    &  $<$24.0        &  1200      &  Membership fraction &  10200  &  4   \\ 
                           &                                                 &                           &                & \& Control field          &                &       \\ 
$R$ (South)     &   -1.28$^{+0.05}_{-0.05}$     &  $<$24.0        &   800         &  Membership fraction  &  10200  &  4   \\ 
                           &                                                 &                           &                  & \& Control field          &                &       \\ 
$I$ (North)        &   -1.60$^{+0.01}_{-0.01}$   &  $<$23.25        &   1200      &  Membership fraction  &  10200  &  4   \\ 
                           &                                                 &                           &                  & \& Control field          &                &       \\ 
$I$ (South)       &   -1.27$^{+0.02}_{-0.02}$   &  $<$23.25        &   800         &  Membership fraction  &  10200  &  4   \\ 
                           &                                                 &                           &                   & \& Control field          &                &       \\ 

\cutinhead{Infrared}

$H$                     &  -1.73$^{+0.14}_{-0.14}$   &  14.0-16.0 (F) &  657          &  $(B-R)$ color  &  \nodata  &  17  \\
$H$                     &  $\sim -1.3$                           &  10.0-17.0  &  480           &  Control field  &  23.7  &  18 \\
$K$                      &  -1.41$^{+0.34}_{-0.37}$    &  15.5-18.5 (F) &   41             & $(B-R)/(B-K)$     & \nodata  &  19  \\
3.6$\mu$m (coma 1)     & -2.18$^{+0.13}_{-0.13}$  & 14.5-17.0 (F) & 2640 & Membership fraction & \nodata   & 20 \\
3.6$\mu$m (coma 3)     & -2.60$^{+0.33}_{-0.33}$  & 14.5-16.5 (F)  & 2000 & Membership fraction &  \nodata  & 20 \\
3.6$\mu$m (total)           & -2.38$^{+0.15}_{-0.15}$  & 14.5-16.5 (F)  & 4640 & Membership fraction &  \nodata   & 20 \\

$24\mu$m (core)  &  -0.99$^{+0.34}_{-0.52}$    &  $<$11  & 452 & Redshift  & \nodata    & 21  \\
$24\mu$m (off-center)  &  -1.32$^{+0.11}_{-0.14}$   & $<$11  &  8680 & Redshift  & \nodata    & 21  \\
$24\mu$m (total)  &  -1.49$^{+0.11}_{-0.11}$   & $<$11   &  14400 & Redshift  &  \nodata   & 21  \\

\enddata
\tablecomments{$^a$ Apparent magnitude range of $\alpha$ slope measurement.  Where only faint data were fitted, or where two-component models were used, the quoted magnitude range and slope is for the faint component only (denoted by ``F").  Otherwise, slopes were measured over the full LF magnitude range. \vspace{-0.7cm} }

\tablerefs{(1) \cite{andreon99}; (2) \cite{cortese03}; (3) \cite{beijersbergen02}; (4) \cite{adami07}; (5) \cite{thompson93}; (6) \cite{biviano95}; (7) \cite{mobasher03}; (8) \cite{andreon02};  (9) \cite{lobo97}; (10) \cite{paramo03}; (11) \cite{paramo02}; (12) \cite{trentham98}; (13) \cite{bernstein95}; (14) \cite{lopez97}; (15) \cite{secker97}; (16) \cite{milne07};  (17) \cite{depropris98}; (18) \cite{andreon00}; (19) \cite{mobasher98}; (20) This work; (21) \cite{bai06}.}
\end{deluxetable*}

\subsection{Membership \& Background Subtraction}
\label{sec:background}

It is important to consider the method of background source subtraction when comparing results, as different methods can yield different values of $\alpha$. Traditionally, three main types of background subtraction have been used: (1) statistical subtraction of control field galaxy counts, (2) color information ($B-R$) to exclude galaxies redder than the average for the cluster and (3) redshift/membership fraction, the method used in this study. Since the advent of wide-field CCD detectors, many studies have used nearby offset control fields for background source subtraction. However unless the control fields are very large (at least the size of a cosmic void), cosmic field-to-field variations in background galaxy counts have the potential to introduce large uncertainties.  This method is required for background subtraction at very faint magnitudes ($R\ga22$) due to current limitations of multi-object spectroscopy.  However, it may not be appropriate for studying a rich cluster with a large angular size like Coma, particularly when one might expect the entire environment (i.e. the region of space behind Coma) to perhaps show over-enhancements in galaxy counts due to background groups, filaments etc.  Indeed, \cite{adami00} detect three galaxy groups ($z\sim$ 0.1-0.3) and a more distant cluster at $z\sim0.5$ along the line-of-sight to the Coma cluster, which would contaminate Coma galaxy counts.  Using simulations,  \cite{valotto01} have also demonstrated that such projection effects, resulting from large-scale structure behind clusters, can produce artificially steep faint-end slopes ($\alpha \la -1.5$).  

In this study, we have spectroscopic information for the entire 3.6$\mu$m magnitude range of the LF, which gives a more reliable estimate of the faint-end slope in our data.  For comparison, however, we have constructed a field-subtracted LF using the field counts of  \cite{fazio04b}, which is shown in Figure~\ref{fig:lf_all} ({\it lower right}).  Although not well constrained, it is consistent in almost all bins with the LF computed using the spectroscopic membership fraction within the combined poissonian errors.  The faint end of the field-subtracted LF will be very uncertain due to the smaller size of the EGS control field (1224 arcmin$^2$), but importantly we demonstrate that the steep upturn is present in the data regardless of the background subtraction method.

\subsection{Comparison to other Coma IR LFs}
\label{sec:compare_ir}

\begin{figure}
\begin{center}
\scalebox{0.42}{\includegraphics{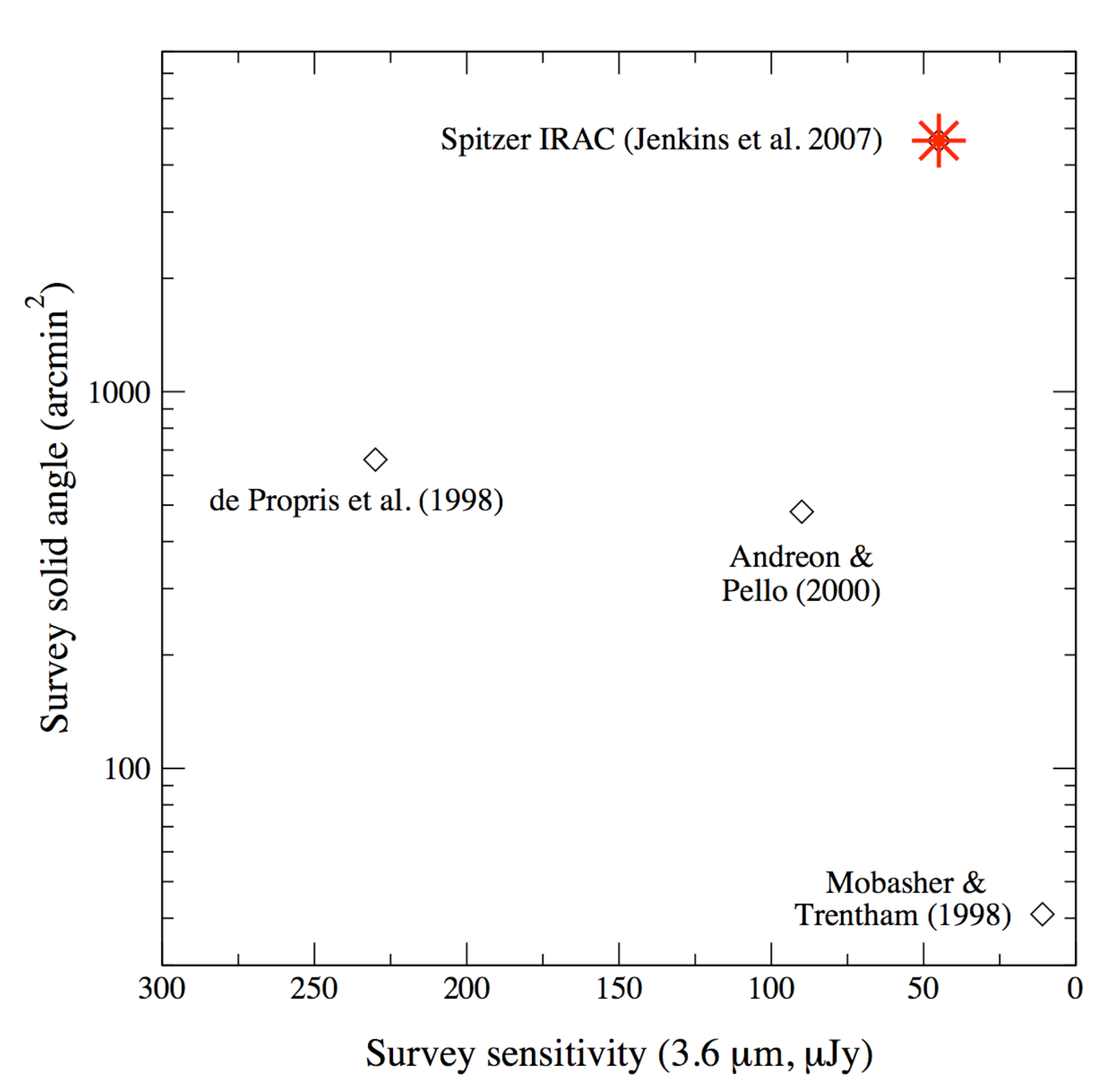}}
\caption{A comparison of previous near-IR coma surveys with the {\it Spitzer} IRAC survey.  All flux densities for the $H$- and $K$-band surveys were transformed to equivalent 3.6$\mu$m flux densities assuming an elliptical SED.  The IRAC survey is $\sim$2 times deeper than the deepest $H$-band survey of \citet{andreon00}, and covers and area 7 times greater than any previous study.}
\label{fig:survey_compare}
\end{center}
\end{figure}

The majority of Coma LF studies have been carried out at optical wavelengths (see \S~\ref{sec:compare_other}). However, our 3.6$\mu$m LFs are more directly comparable with near-IR LFs at the slightly shorter wavelengths of $H$ (1.6$\mu$m) and $K$ (2.2$\mu$m) as they also probe the stellar mass of the galaxies.  Figure~\ref{fig:survey_compare} summarizes the properties of this and previous near-IR surveys, and demonstrates that the IRAC survey covers a previously unexplored regime of a large surface area combined with good photometric depth.  

The first $K$-band Coma LF was measured by \cite{mobasher98} for a small 41 arcmin$^2$ area in the core down to $K = 18.5$.  This LF was not well constrained ($\alpha=-1.41^{+0.34}_{-0.37}$ with M$^\star_K$ fixed at -24.35) due to large uncertainties in the background source subtraction, but the faint galaxies were shown to be dwarf spheroidals by their optical-IR colors.  A better comparison comes from the larger area (660 arcmin$^2$) $H$-band study of the core region by \cite{depropris98}, where the shape of the LF (down to $H\sim16$) is similar in many ways to the 3.6$\mu$m LF presented here.  The statistics ($\sim$200 galaxies compared to only $\sim$ 40 in the \citealt{mobasher98} study) allowed better constraints on the parameters, and the bright end was fitted with a Schechter function with $\alpha=-0.78$ and M$^\star_H=-23.87$ (for DM = 35.0). Note that \cite{depropris98} also found evidence of an excess of galaxies fainter than $H=14.5$ mag; using two different background subtraction methods (statistical field subtraction and $B-R$ color selection), they measured a power-law slope of $\alpha=-1.7$.  Another $H$-band study by \cite{andreon00} covered a smaller area (480 arcmin$^2$) $\sim$15\arcmin\ North-East from the cluster center, and reached one magnitude deeper ($H\sim17$) than \cite{depropris98}.  This $H$-band LF was fitted with a single Schechter function with a slope of $\alpha\sim-1.3$ and M$^{\star}_H = -23.9$ (for DM = 35.0), but again was not well-constrained.   A dip was also seen at M$_{H}\sim-22$, which is also present in both fields in the 3.6$\mu$m LFs and in the \cite{depropris98} $H$-band data.  We also see another shallow dip in the IRAC data in both fields at M$_{3.6\mu m}\sim-23.75$.   These dips have been seen in optical surveys of Coma (\citealt{godwin83}; \citealt{biviano95}; \citealt{mobasher03}; \citealt{adami07}), and are thought to be a result of the merging and disappearance of bright elliptical galaxies (\citealt{andreon00}; \citealt{adami07}).  

All of these near-IR LFs have relatively flat faint-end slopes, similar to the IRAC 3.6$\mu$m LF when fitted by a Schechter function over the range $-27 <$ M$_{3.6\mu m} < -20.5$.  However, apart from the  \cite{depropris98} $H$-band study, they do not show evidence of the excess population of faint (M$_{3.6\mu m} > -20.5$) galaxies seen here.  This is likely to be a direct consequence of the increased sensitivity of IRAC (i.e. darker sky), permitting much better statistics compared to the ground-based instruments (see \S~\ref{sec:intro}). The vastly lower sky background level at 3.6$\mu$m will also increase detection of low-surface-brightness (LSB) dwarf galaxies that may have been missed in the ground-based surveys.

Recently, there has also been a longer-wavelength {\it Spitzer} Multiband Imaging Photometer (MIPS) 24$\mu$m study of the Coma LF \citep{bai06}. However, due to the larger PSF of MIPS (6\arcsec\ FWHM), it is much more prone to confusion issues in crowded fields like Coma and cannot therefore go as deep as IRAC.  \cite{bai06} detected 217 member galaxies at 24$\mu$m, but only down to m$_ {24\mu m}=11$ (equivalent to m$_ {3.6\mu m}\sim$15 assuming an elliptical SED).  Since the MIPS data were not deep enough to reach the dwarf galaxy population turn-up seen in the IRAC data, only single Schechter functions were required to fit the data, with slopes of $\alpha = -0.99^{+0.34}_{-0.52}$ in the core and $\alpha = -1.32^{+0.11}_{-0.14}$ in their off-center region. These results are consistent with the 3.6$\mu$m bright-end slopes in the two fields, within the combined errors.

\subsection{Comparison with Field LFs}
\label{sec:compare_field}

We can also compare the space density of 3.6$\mu$m detected galaxies in the Coma LF to that of the field population over the same magnitude range.  As shown in \S~\ref{sec:fieldcounts}, in addition to the obvious excess at bright magnitudes, there is also a significant excess of galaxies in Coma compared with 3.6$\mu$m field galaxy counts of \cite{fazio04b} at faint magnitudes (m$_{3.6\mu m}>14.5$), the position of the Coma upturn (see Figure~\ref{fig:galcounts}).  

Recent determinations of field $K$-band LFs are deep enough to compare with Coma (e.g. \citealt{bolzonella02}; \citealt{pozzetti03}; \citealt{feulner03}; \citealt{saracco06}), and they are all consistent with a flat faint-end slope for the LF ($\alpha\sim$ -1.0 to -1.3) out to redshifts of $z < 1.3$.   A 3.6$\mu$m field LF has also been published by \cite{babbedge06} for the {\it Spitzer} SWIRE survey, which is deeper than our Coma survey (m$_ {3.6\mu m}\sim$19.5).  This also shows a flat faint-end slope, with $\alpha\sim$ -0.9 to -1.0 over the redshift range $z\sim$ 0 to 1.5.  The excess of galaxies detected in Coma with respect to 3.6$\mu$m field galaxy counts, together with relatively flat slopes of near-IR field LFs extending beyond the depth of our IRAC Coma sample, support the evidence that the large dwarf population detected at 3.6$\mu$m is not an artifact of incorrect background source subtraction, and that they do indeed belong to Coma.

\subsection{Comparisons with Coma LFs at other wavelengths}
\label{sec:compare_other}

\begin{figure*}
\begin{center}
\scalebox{0.7}{\includegraphics{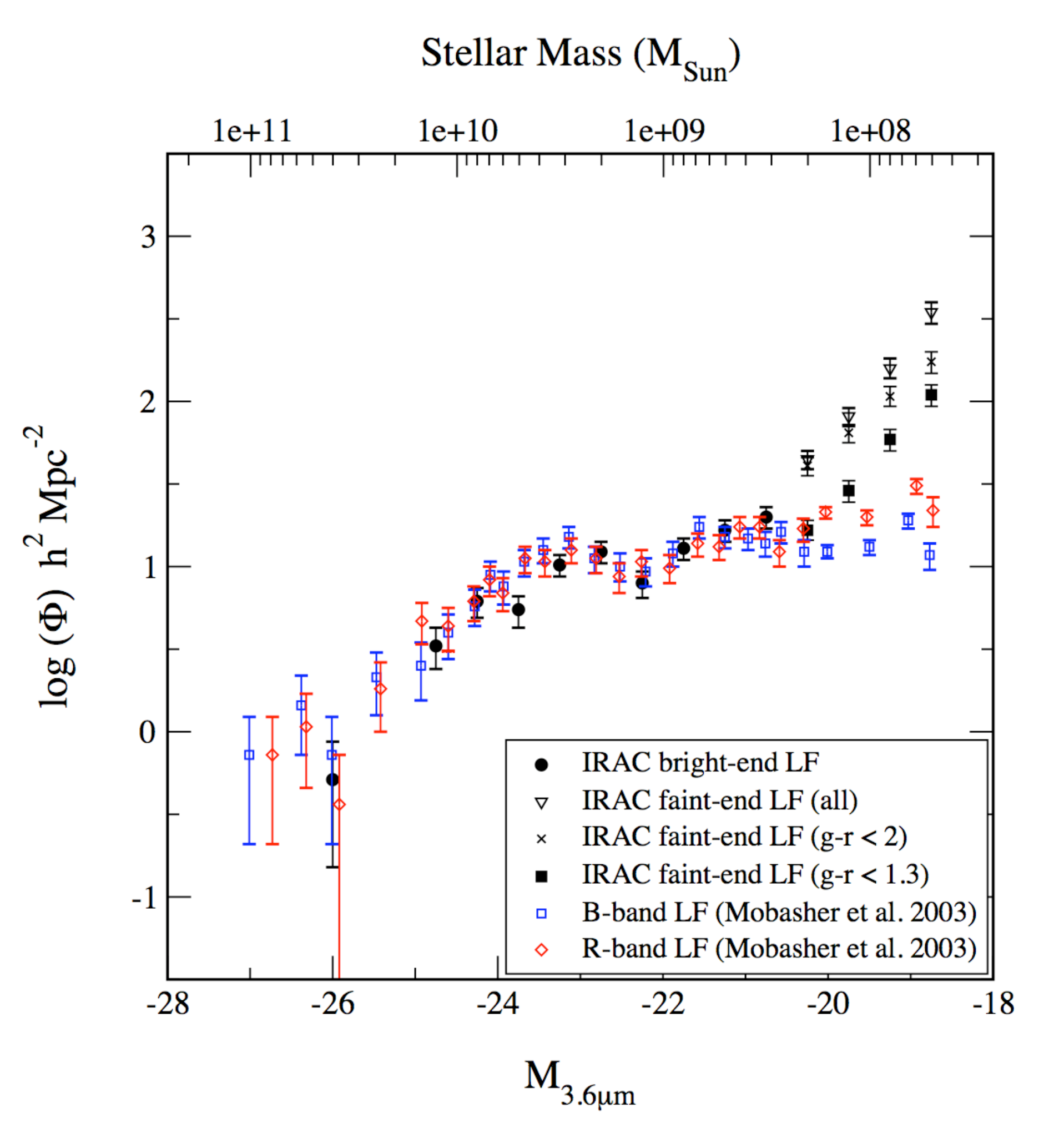}}
\caption{A qualitative comparison of the combined 3.6$\mu$m LF (Coma 1 + Coma 3) with the equivalent $R$- and $B$-band optical LFs of \cite{mobasher03}.  These are directly comparable to the IRAC LF as the same method of background source subtraction was used (i.e. spectroscopic membership fraction).  The optical LFs have been shifted to the 3.6$\mu$m frame using the mean optical-IR colors of the confirmed cluster members per half-magnitude bin.  The range of stellar masses covered by these observations is also shown, assuming a $K$-band mass-to-light ratio of 0.91 at $z=0$ \citep{drory04}.  The 3.6$\mu$m and optical LFs agree well at the bright end up to M$_{3.6\mu m}\sim -20.5$.  At this point, equivalent to M$_{R}\ga-18.0$, there is a steep rise in the IR LF towards fainter magnitudes compared to the optical, presumably with the onset of the dwarf galaxy population. Note that while one can compare the {\it shapes} of the optical and IRAC LFs, a comparison of their relative normalizations may be misleading; the excess of IR-detected galaxies may represent a LSB dwarf population, or a population too red to be detected down to the optical limit of $R\sim20$ (or a combination of both).}
\label{fig:lf_compare}
\end{center}
\end{figure*}

The optical LF studies of Coma are deep enough to reach the dwarf galaxy population, and a wide range of faint-end slopes have been measured ($\alpha\sim$ -1.0 to -1.8, see Table~\ref{tab:lfsummary}, e.g. \citealt{thompson93}; \citealt{bernstein95}; \citealt{biviano95}; \citealt{lopez97};  \citealt{secker97};  \citealt{beijersbergen02}; \citealt{andreon02}; \citealt{lobo97}; \citealt{trentham98}) .  A more recent HST $R$-band study measures an even steeper slope of -2.3 in a small 5 arcmin$^2$ core region \citep{milne07}.

We can {\it qualitatively} compare the 3.6$\mu$m LF to the $R$ and $B$-band optical LFs of \cite{mobasher03}, as the background subtraction was also performed using spectroscopic membership fractions.   To translate the optical LFs to their expected magnitudes at 3.6$\mu$m, we compute mean optical-IR colors of the spectroscopically confirmed members to be $R$-3.6$\mu$m $\sim$ 2.7 and $B$-3.6$\mu$m $\sim$ 4.2 for all galaxies (without regard to galaxy classification).  However, due to the known luminosity/galaxy type segregation in clusters, the corrections will not be uniform across all magnitude bins.  In order to match the optical and 3.6$\mu$m LFs more accurately, we derive the mean $R-3.6\mu$m  and $B-3.6\mu$m colors of the confirmed members in half-magnitude bins, corresponding to the binning of the optical and IR data.  

Figure~\ref{fig:lf_compare} shows the combined (Coma 1 + Coma 3) 3.6$\mu$m LF, together with the equivalent $R$- and $B$-band LFs of \cite{mobasher03} shifted by the mean $R-3.6\mu$m and $B-3.6\mu$m colors.  Also shown is the range of stellar masses covered by the IRAC observations, which are estimated as $M_{\ast} (M_{\odot}) \sim 5.7\times10^5 f_{3.6\mu m}$, where $f_{3.6\mu m}$ is the 3.6$\mu$m flux density in $\mu$Jy.  This is calculated using a distance of 100\,Mpc and a $K$-band mass-to-light ratio of 0.91 at $z=0$ \citep{drory04}, after transforming the 3.6$\mu$m flux densities to $K$-band using the {\it Spitzer} EX-PET\footnote{http://ssc.spitzer.caltech.edu/tools/expet/} tool, assuming an elliptical SED.

\begin{figure}
\begin{center}
\scalebox{0.45}{\includegraphics{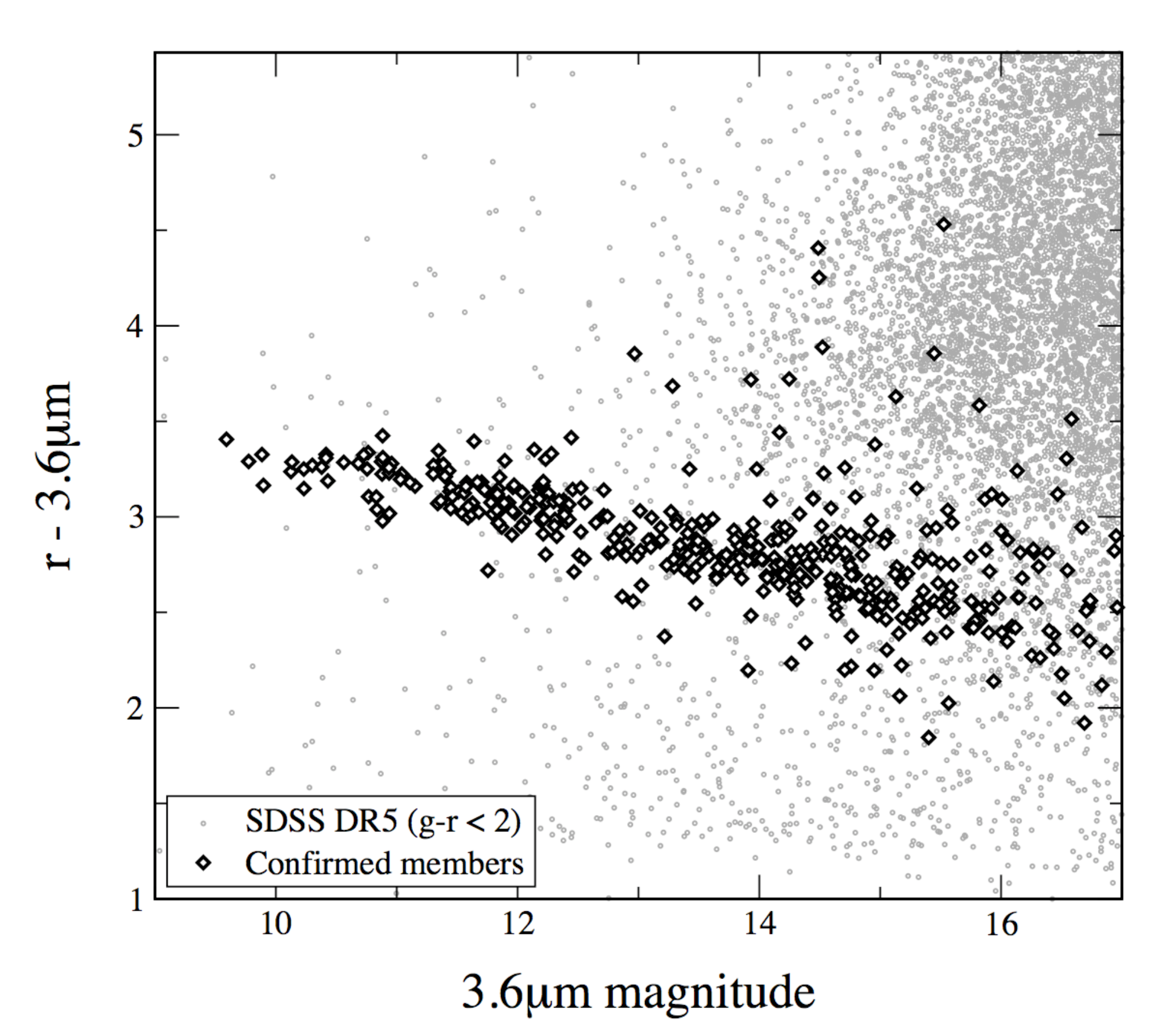}}
\caption{The $R-3.6\mu$m versus 3.6$\mu$m color-magnitude relation for the $r<22.5$, $g-r<2$ color-selected subset of IRAC-detected galaxies matched to the SDSS DR5 catalog.   A strong red sequence is formed by the Coma members at $R-3.6\mu$m~$\sim2.5-3.5$, but note that there is an increase in redder members towards fainter magnitudes.  There is also a general reddening of all galaxies at faint magnitudes.  Although a fraction of these will be background galaxies, this is consistent with the suggestion that the large IRAC-detected faint population in Coma may be redder than the optically-selected faint population.}
\label{fig:opt_ir}
\end{center}
\end{figure}

The bright-ends of the LFs agree well, which is to be expected as we are sampling almost the same galaxy populations.  However, fainter than M$_{3.6\mu m}\sim$-20.5, the shapes differ significantly, with the IRAC LF having a much steeper slope than the optical LFs.   At the faint end, we show both the raw 3.6$\mu$m LF (with no optical selection) and the two $g-r$ color-cut LFs, illustrating that the IR-excess is still significant with these constraints.  It is, however, important to note that while one can compare the {\it shapes} of the optical and IRAC LFs, a comparison of their relative normalizations may be misleading.  A possible explanation for the difference between the optical and IR LFs is that the apparent excess of faint IR-detected galaxies may represent either a LSB dwarf population that IRAC is more sensitive to, or a population of dwarf galaxies that are too red to be detected in \cite{mobasher03} down to the optical limit of $R\sim20$ (or a combination of both).  

However, there are many other optical surveys that probe much deeper into the dwarf population, which, by necessity, use statistical background source subtraction (e.g. \citealt{bernstein95} ($R<23.5$), \citealt{trentham98} ($R<24$) and \citealt{adami07} ($R<24$)). The optical color-cuts we impose on our 3.6$\mu$m faint-end LFs means that we do not include any optically undetected galaxies.  However, the SDSS catalog only reaches $r\sim22.5$, so in terms of $r$-band magnitude, this cut may be overly-conservative.  There could be additional IRAC-detected dwarfs that are optically fainter than this, and indeed the deeper optical studies already show evidence of such faint populations. For example, the $R$-band membership fraction of \cite{mobasher03} implies a flat LF slope out to $R\sim20$, but the deeper $R$-band LFs of \cite{bernstein95}, \cite{trentham98} and \cite{adami07} all continue to rise beyond this with slopes of $\alpha\sim$ -1.4 to -1.7.  This may therefore be the $R$-band regime where the excess 3.6$\mu$m dwarf galaxy population is detectable, and suggests that this near-IR survey may manifest this population.  To investigate this further, we have cross-correlated the Coma 1 IRAC catalog with the deep \cite{adami07} optical catalog over an area of 1365 arcmin$^2$ in the central region common to both surveys.  We find that $\sim$93\% of the faint 3.6$\mu$m sources (m$_{3.6\mu m}>14.5$) have optical matches down to $R=24$. 

These results imply that the faint IR population has much redder optical/IR colors compared to the optically selected dwarfs in the cluster.  Since we have used membership fractions to construct the LF, we have not yet identified which {\it particular} galaxies in this population are Coma members.   However, in Figure~\ref{fig:opt_ir} we show the $r-3.6\mu$m versus 3.6$\mu$m color-magnitude relation for the $r<22.5$, $g-r<2$ optically-selected subset of IRAC-detected galaxies matched to the SDSS catalog.  In a similar fashion to the optical color diagram (Figure~\ref{fig:sdss}), the Coma members form a strong red sequence at $r-3.6\mu$m~$\sim2.5-3.5$, but we also note an increase in redder members towards fainter magnitudes.  There is also a general reddening of all galaxies at faint magnitudes; although a fraction of these will be background galaxies, this is consistent with the suggestion that the IRAC-detected faint population in Coma is redder than the optically-selected dwarf population.  This will be explored in more detail in a future paper.

\subsection{Comparison of faint-end slope versus wavelength}

\begin{figure}
\begin{center}
\scalebox{0.48}{\includegraphics{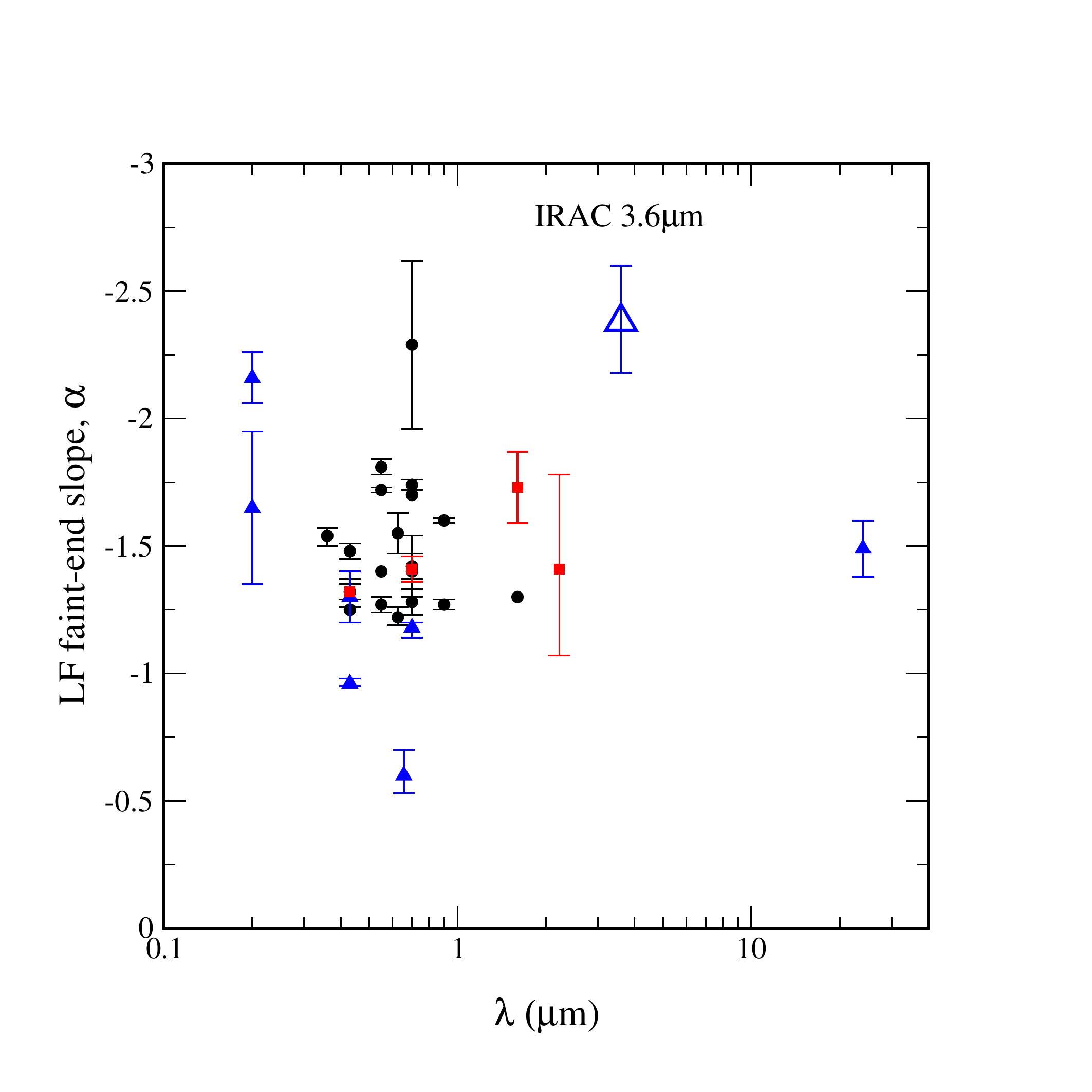}}
\caption{Measured values of the Coma LF faint-end slope $\alpha$ from the literature as a function of wavelength (see Table~\ref{tab:lfsummary}).  The symbols denote different methods of background source subtraction:  control field ({\it black circles}), $B$-$R$ color ({\it red squares}) and redshifts/membership fractions ({\it blue triangles}).  The 3.6$\mu$m `raw' faint-end slope (with no optical color-cut) is highlighted with a large blue open triangle, with an error bar representing the full range of $\alpha_2$ obtained with all $g-r$ color cuts in both Coma fields.  No significant systematic variation of slope with wavelength is evident in these combined results. The 3.6$\mu$m slope is steeper than all but one measurement in the UV \citep{andreon99} and the HST $R$-band core measurement of  \citep{milne07}.}
\label{fig:alpha_compare}
\end{center}
\end{figure}

Figure~\ref{fig:alpha_compare} shows the multi-wavelength Coma faint-end LF slopes from Table~\ref{tab:lfsummary} plotted versus filter wavelength.  Although the areas and photometric depths covered by these studies vary, they can be used to investigate any broad trends in changes of the faint-end slope with wavelength.  Where slopes were measured in different regions of Coma in a particular study (i.e. core/off-center/total), only the slope of the total area LF is plotted.    Some evidence has previously been found for change of the faint slope between different bands (e.g. \citealt{beijersbergen02}; \citealt{adami07}), but no significant trend is evident in this figure.   The different symbols/colors denote the different methods of background source subtraction used.  At optical wavelengths, the slopes based on samples measured with a spectroscopic background subtraction appear systematically shallower (in agreement with e.g.  \citealt{valotto01}, \citealt{adami00}).  However, the UV measurement of \cite{andreon99} and \cite{cortese03}, both using a similar method, are much steeper ($\alpha$ = -2.16 and -1.65 respectively).  The 24$\mu$m slope measured by \cite{bai06} using redshift membership fractions ($\alpha=-1.49$) is consistent within the errors with the optical measurements using statistical background source subtraction.  However,  this is relatively shallower and therefore not directly comparable.  A systematic study using reliable redshift background source subtraction across all wavebands is required to properly determine if there is any color evolution of the faint-end slope.  Note that the IRAC 3.6$\mu$m slope is significantly steeper than all but one measurement in the UV \citep{andreon99} and the deep HST $R$-band study of \cite{milne07}.

\subsection{Environmental Dependence}
\label{sec:environ}

One of the goals of this study is to search for changes in the shape of the faint end of the LF in different regions of Coma. The shape of the LF is not thought to be universal, but strongly influenced by local environment.  For instance, \cite{lopez97} measure flat slopes ($\alpha\sim-1.0$) in a sample of rich clusters, and find much steeper slopes in poorer clusters ($-2.0 < \alpha < -1.4$).  In the central regions of clusters, processes such as tidal stripping and ram pressure stripping \citep{gunn72} conspire to destroy smaller dwarf galaxies by removing their stars and gas.  Another destructive process that depends on cluster density is galaxy harassment \citep{moore98}, where frequent and rapid close encounters of small spiral galaxies convert them into dwarf spheroidal systems. In addition, \cite{adami00} suggest that the high temperatures present in Coma during its initial formation could have suppressed the formation of dwarf galaxies in the core region.  These effects point towards a scenario where the number of dwarf galaxies should decrease towards the center of the cluster, leading to a flattening of the LF faint-end slope.   However there are other competing factors that come into play.  The denser the environment, the more likely that dark halos of galaxies could collect gas from the inter-cluster medium to be converted into stars, leading to larger numbers of dwarf galaxies in denser evolved cores than the less-dense, unevolved off-center or field (\citealt{mobasher03} and references therein).  This could counter-balance the effects of destructive forces of tidal interactions, ram pressure stripping and harassment.

We do find significant differences between slopes at the core and off-center regions of the cluster. If this directly traces the mass, then a difference in the LFs means galaxies have different formation/destruction and merger histories depending on their local environment.  At the bright end (Schechter form), the slope increases from $\alpha_1$ = -1.18$\pm$0.06 in the core to -1.30$\pm$0.08 in the off-center field, while at the faint end (power-law form) there is an increase from $\alpha_2$ = -2.18$\pm$0.13 to -2.60$\pm$0.33 (see Table~\ref{tab:fits} and Figure~\ref{fig:lf_all}). Additionally,  M$^{\star}_{3.6\mu m}$ is fainter in Coma 3, which reflects the fact that this off-center region does not possess as many of the very bright galaxies seen in the central region. Some previous studies covering sufficiently large areas of Coma have also found similar changes in LF slope with environment, as shown in Table~\ref{tab:lfsummary}.  For example, \cite{beijersbergen02} finds a $U$-band slope of -1.32 in the core region, but a steeper slope of -1.54 when all areas of the cluster within the survey are included.  A smaller increase is also found in the $r$-band, with no such difference found in the $B$-band.  A similar effect is found in the $V$-band by \cite{lobo97}.   More recently, \cite{adami07} have found significant differences in the faint-end slopes of northern and southern areas of the center of the cluster, with a steep $R$-band slope in the north (-1.74) and a flatter slope in the south (-1.28).  \cite{mobasher03} also find a small increase in the $R$-band slope between the core (-1.17) and the off-center (-1.29), and  \cite{bai06} found a similar effect at mid-IR wavelengths, with a slope of -0.99 in the core, increasing to -1.32 in their off-center region.  Collectively, these results provide strong evidence that dwarf galaxies are being destroyed or merged into larger galaxies in the denser central regions of the cluster.  This scenario is further supported by evidence of disruption in the central regions of Coma in the form of tidal features, resulting from galaxy-galaxy and galaxy-cluster interactions (\citealt{trenthammobasher98};  \citealt{gregg98}).

\subsection{Redshift evolution of the IR LF}

It is interesting at this stage to compare the Coma IR LF with examples at higher redshifts to study possible evolutionary trends.  In \cite{depropris99}, the $K$-band LFs of 38 clusters with redshifts in the range $0.1 < z < 1$ were measured.  Although they were unable to constrain the faint-end slopes of the clusters with these data, they found that the evolution of M$^{\star}_K$ was consistent with a model where the giant early-type galaxies in the clusters formed all their stars in a single burst at high redshift ($z > 2$), and evolved passively thereafter.  This result was independent of the richness of the clusters.

More recently, there have been two studies of the IRAC 3.6$\mu$m redshift evolution of clusters.  \cite{muzzin05} measured the 3.6$\mu$m LFs of 123 galaxy clusters in the {\it Spitzer} First Look Survey (FLS) in the redshift range $0.15 < z <1.22$.  They found that M$^{\star}_{3.6\mu m}$ evolved in a manner consistent with a passive evolution model (in agreement with \citealt{depropris99}). Moreover, they also found some evidence that the faint-end slope became shallower at $z > 0.5$.  Together, these results imply similar redshifts ($z > 2$) for the formation of massive galaxies and clusters, with many of the dwarf galaxies accreted from the field at $z < 0.5$.  This result is also supported by \cite{toft04}, whose $K_S$-band LF of a cluster at $z=1.237$ has a faint-end slope of $\alpha=-0.64$, which is significantly shallower than clusters in the local Universe.  Another study by \cite{andreon06} of 32 clusters in the {\it Spitzer} SWIRE survey in the redshift range $0.2 < z < 1.25$ also found that their values of M$^{\star}_{3.6\mu m}$ were consistent with a passive evolution model, but were unable to determine any change of slope with redshift with these data.

The large population of dwarf galaxies detected at 3.6$\mu$m in Coma supports the scenario of accretion of dwarf galaxies from the field at low redshift.  However, our values of M$^{\star}_{3.6\mu m}$  (-25.17$\pm$0.10 for Coma 1 and -24.69$\pm$0.15 for Coma 3) are slightly brighter than the value predicted by the passive evolution model of \cite{bruzual93} of -24.33 (A. Muzzin, {\it private communication}).

\section{Summary \& Conclusions}
\label{sec:conc}

We have conducted a wide-field {\it Spitzer} IRAC 3.6$\mu$m survey of two fields in the Coma cluster, in which we have dramatically improved constraints on the faint end of the near-IR Coma LF compared to previous IR surveys.  Thanks to the efficiency of the IRAC instrument and the significantly reduced sky background in space, we have been able to probe as faint as m$_{3.6\mu m}\sim17$ (M$_{3.6\mu m}\sim-18$) over a large fraction of the cluster ($\sim1.3$ deg$^2$), giving a very representative view of the IR dwarf population of Coma.  The first of our two fields covers an area of 0.733 deg$^2$ in the core of the cluster (Coma 1), and the second is an off-center 0.555 deg$^2$ region,  $\sim$57\arcmin~ from the core (Coma 3).   We detect a total of 29,208 sources, 17,872 in Coma 1 and 11,336 in Coma 3.  Our main results can be summarized as follows:

\begin{enumerate}
 
\item We construct 3.6$\mu$m LFs for the Coma 1 and Coma 3 fields. To subtract the contribution to the 3.6$\mu$m counts due to background galaxies, we use spectroscopic redshifts for IRAC-detected galaxies to construct spectroscopic membership fractions for each field.  At the bright end (M$_{3.6\mu m}<-20.5$), the LFs can be fitted with a classic Schechter function with slopes of $\alpha_1$ = -1.18$\pm$0.06 (Coma 1) and -1.30$\pm$0.08 (Coma 3), and M$^{\star}_{3.6\mu m}$ values of -25.17$\pm$0.10 and -~24.69$\pm$0.15 for Coma 1 and 3 respectively.  

\item Faintward of M$_{3.6\mu m}\sim-20.5$, {\it we detect a steep faint-end slope in both fields}. To guard against any optical selection bias in the spectroscopic completeness functions, we use SDSS photometric data to construct optically-selected 3.6$\mu$m faint-end slopes for both fields.  Taking only the IRAC sources with optical counterparts at $r<22.5$ mag and cut at $g-r<1.3$, we measure steep faint-end slopes of $\alpha_2$ = -2.18$\pm$0.13 in Coma 1 and -2.60$\pm$0.33 in Coma 3, which infer that there are at least $\sim1600$ member galaxies over the two fields.  The differences in slopes between the two fields provides evidence for environmental effects. 

\item When qualitatively compared to $B$- and $R$-band optical LFs constructed from the same areas of Coma and using the same spectroscopic completeness method, we find that the optical LFs have fairly flat distributions at the faint end ($\alpha\sim-1.2$).   A possible explanation for the difference between the optical and IR LFs is that the excess of faint IR-detected galaxies may represent either a LSB dwarf population that IRAC is more sensitive to, or a population of dwarf galaxies that are too red to be detected in the optical survey down to the limit of $R\sim20$ (or a combination of both).

\end{enumerate}

\noindent This is the first {\it Spitzer} IRAC 3.6$\mu$m study of a low-redshift cluster, and it has demonstrated that IRACs combination of sensitivity, spatial resolution and mapping capabilities makes it the current instrument of choice to probe the stellar mass distribution of clusters.  The steep faint-end rise seen at 3.6$\mu$m has huge implications towards constraining the mass function of galaxies, and provide clues for an optically faint population of low mass galaxies in clusters.  Approximately $60$\% of the IRAC sources have optical counterparts in the SDSS data down to $r\sim22.5$, and $>$ 90\% of a subset in the Coma 1 field have counterparts down to $R=24$.    There is therefore great potential that the numbers of IRAC detected galaxies implied by our SDSS optical color-cuts as belonging to Coma will increase with further investigation.  

The chief source of uncertainty at faint magnitudes in studies such as this is spectroscopic completeness, which for Coma is $<$ 10\% at $r=21$. To overcome this, we are conducting wide-field spectroscopic programs to dramatically increase the coverage of faint sources, which will give us a much better constraint on the contribution to the background from field galaxies and background clusters/groups.  However, to cover thousands of objects over square degree fields of view (as is needed in Coma) is still observationally very difficult.  The current state-of-the-art in wide-field multi-object spectroscopy reaches $r\sim22$ in several hour exposures (e.g.  the Hectospec work of \citealt{papovich06}), and so at present, statistical field subtraction must still be used fainter than this.  

We are also conducting a multi-wavelength study of the faint galaxies detected by IRAC, to measure their photometric redshifts (and hence their Coma membership), and also to ascertain properties such as stellar ages, star formation histories and morphological types.  Other studies using new GALEX UV and {\it XMM-Newton} X-ray data of these Coma fields are also in progress.  Through such multi-wavelength characterizations of the galaxy populations of Coma, we will gain a better understanding of galaxy evolution in cluster environments.

\begin{center}  
ACKNOWLEDGMENTS
\end{center}

\noindent We thank the referee for helpful comments.  We also thank N. Trentham, R. Marzke and H. Ferguson for helpful discussions,  and R. Arendt for supplying the DIRBE Faint Source Model for Coma.  This research was partly funded by the {\it Spitzer} Science Center for observing program \#3521 (P.I. AEH), based on observations made with the {\it Spitzer Space Telescope}, which is operated by the Jet Propulsion Laboratory, California Institute of Technology, under NASA contract 1407. LPJ also acknowledges funding from the NASA Postdoctoral Fellowship Program.   DMA acknowledges the Royal Society for support.

\end{document}